\documentclass[manuscript]{aastex}

\newif\ifAMStwofonts
\newcommand{\be}{\begin{equation}}
\newcommand{\ee}{\end{equation}}
\newcommand{\bea}{\begin{eqnarray}}
\newcommand{\eea}{\end{eqnarray}}

\begin{document}


\author{R.N. Henriksen}

\affil{Physics, Engineering Physics \& Astronomy,Queen's University} 
\affil{Kingston, Ontario, K7L 2T3, Canada}
\affil{henriksn@astro.queensu.ca}
\date{\today}



\begin{abstract}

We study transient spiral structures in an isothermal, thin, galactic  disc.  We find no instability that can grow everywhere from infinitesimal disturbances, but spiral structure does grow in the disc due to an arbitrarily strong, asymmetric, central source. An initially finite spiral undergoes transient swing amplification as it is gradually wound-up by differential rotation. An independent sequence in negative time describes a leading spiral swinging to a trailing spiral. The dynamical coupling  is established between the swinging potential and the arm particles, by ensuring that this potential constrains a locally rotating distribution function centred on the arms. This swing amplification propagates in radius at the constant rotational speed of the disc, and leaves behind an exponential density decline in space and time.     
 
     
\end{abstract} 
\keywords{galaxies-spiral, galaxies-structure, gravitation, spiral arms}
\title{Swinging Spiral Arms}
\setlength{\baselineskip}{13pt}
\section{Introduction}
\label{sec:intro}

In this paper we study  logarithmic spiral structure in an isothermal or `Mestel' disc. The axi-symmetric disc is necessarily present as is the axi-symmetric halo, but we ignore the non-axially symmetric halo for clarity. It generally has a higher order effect on the disc structure. Non-axially symmetric halo structure, consistent with the disc spiral structure, has been studied in a paper that is available on the Arxiv (\cite{Harxiv2012}).  

The spiral structure that we describe in this paper propagates in radius and azimuth, and so might be described as a spiral `wave'. However it is not a propagating wave in the sense of the well known linear theory (\cite{LS66}: see \cite{BT08} for a description of later developments). Rather, the non-axially symmetric collisionless disc particles are co-moving in the  mean with the spiral potential.  Recent simulations, (\cite{WBS11}), (\cite{KGC11}), and observations (\cite{FRDLW11}) encourage this point of view. It is this co-motion of the particles with the moving arm that leads us to infer an analogy to the classical `swing amplification' (\cite{T82}). 

The evolution of the finite spiral potential is in terms of a variable $\zeta\equiv Vt/r$, where $V$ is the constant disc velocity and $t,~r$ have their usual temporal and radial meanings. Consequently recurrent structure in $\zeta$ generates recurrent structure in time and space. We find that a given initial spiral evolves through a steep transient density amplification near a particular value of $\zeta$, as it swings and winds up. Subsequently this amplified structure dissipates exponentially as $\zeta$ increases. Consequently the recurrence of similar structure persists only for a finite interval or range of radius. Afterwards the process must be re-started either by external disturbance or by some internal linear instability (`self-excitation') if the disc is isolated. 

Such self-excitation behaviour is well known from extensive simulations (e.g. \cite{JS2011}) of isolated discs, even in the tapered Mestel disc that is predicted to be linearly stable (\cite{Z76}), (\cite{ER98b}) except for $m=1$. However the full Mestel disc, which is the self-similar case, is not expected to have well-posed normal modes (\cite{GE99}) in the absence of inner and outer boundaries. It is natural then to expect the full Mestel disc dynamics to depend only on the self-similar variable $\zeta$, since this reflects locally the absence of boundaries.  

In agreement with prior expectations we do not find a truly self-excited instability originating from infinitesimal perturbations. There is instead an indefinitely growing spiral structure in the disc in terms of the variable $\zeta$ that originates from the singular centre of the system. We infer this because although the disturbance is zero everywhere at $t=0=\zeta$, it becomes infinitely strong at the origin for any $t>0$ as $\zeta\rightarrow\infty$. This singular centre should be replaced by a realistic central spiral in a centrally tapered Mestel disc in order to study it as the  origin of  disc spiral instability. This calculation has not been done in this work.

  Despite this central singularity there may nevertheless be an indication of a genuine  instability in the full Mestel disc, if the spiral wave may be regarded as reflected at large radius (where $\zeta=0$). This is implied by the existence of a complementary solution that can be interpreted as due to spirals that have passed from trailing to leading after passage through the centre. However in this paper we do not pursue this possibility, and rather focus on the evolution of an initially given spiral arm.

 We may understand an initially given spiral arm as the ultimate  development either of an externally produced disturbance or of some self-excitation. That is, the arm is comprised of particles that have been entrained by the wave and subsequently move collectively with it. As remarked above such an emergent  wave has been found to arise from an unexpected  instability in recent simulations (\cite{JS2012}) of an isolated Mestel disc. It is suggested in that work that the instability arises from reflection of the wave at its inner Lindblad resonance, where the distribution function (DF) is modified. 
Our self-similar structure traps the particles into a narrow velocity distribution about local co-motion, so that the linear Lindblad resonances do not exist. These are replaced by the extended particle trapping in the winding non-linear potential, which is the reason for the swing amplification. Thus there is no `cavity' essential to the amplification studied here.

The DF of the trapped particles is found from the Collisionless Boltzmann Equation (CBE), as written in the locally co-rotating frame established by the axi-symmetric disc and halo. It is required to share the isothermal self-similarity of the Mestel disc, and it is necessarily time dependent due to the differential rotation. The trapping potential must also take a specific isothermal self-similar form.

The details of the preceding assertions are the subject of the following sections.
In the next section we introduce the formalism for a compatible co-moving, non-axially symmetric isothermal DF and potential. This structure is freely falling in an axi-symmetric background disc and halo. In a following section the coupling between the spiral potential and the comoving DF of the trapped particles is deduced, and the potential solution is found. Finally the spiral wave evolution is analyzed graphically and analytically.   

\section{Transient, Corotating, Spiral Structure in the  Disc}

In this section we construct spiral potential/surface density in a background isothermal disc. Such a disc  rotates with a constant mean circular speed  $v_\phi=V$. We know that this can not be a steady configuration because non axially symmetric structure  winds up  in time due to the differential rotation. We  treat the subsequent time dependence explicitly. We find first a distribution function (DF) for the collisionless particles that are comoving with the spiral arms and then give a key equation for the compatible potential.

We treat the spiral structure by remaining close to an isothermal self-similar evolution in time, at least before major winding has occurred. This is consistent with the initial self-similarity imposed by the Mestel disc. For the collisionless particles the explicit collisionless Boltzmann equation (CBE) equation is written in a differentially rotating frame with angular velocity $\Omega=V/r$. This places our analysis in a reference frame that is in free fall in the axi-symmetric gravitational background due to the disc and halo.  The equation becomes (\cite{Harxiv2012})

\bea
&\partial_tF&+v_r\partial_r F+(\frac{v_\phi}{r}-tv_r\partial_r\Omega)\partial_\phi F\nonumber\\
&+&\!\!\!(\frac{v_\phi^2}{r}+2\Omega v_\phi+\Omega^2 r-\partial_r\Phi_{dr})\partial_{v_r}F\nonumber\\
&-&\!\!\!(\frac{v_\phi v_r}{r}+2\Omega v_r+v_r r\partial_r\Omega+\frac{1}{r}\partial_\phi\Phi_{dr})\partial_{v_\phi}F=0,\label{eq:stac2be}
\eea  
where $F(r,\phi,v_r,v_\phi;t)$ is the two dimensional DF. Other quantities have their usual meaning except $\Phi_{dr}$, which designates the  non-axially symmetric potential of the disc, labelled `disc-rotating'.

The formal  self-similar procedure has been discussed  elsewhere (\cite{LeDHM11b}, and references therein) so we will only outline it here. The self-similarity is constrained to be isothermal by the existence of the constant rotational velocity $V$.
We use a logarithmic time $T$ as the self-similar Lie parameter and introduce on dimensional grounds  the scaled quantities $R$, $\vec{Y}$, $\xi$, $\Psi$ and $P$ according to  
\bea
&\alpha t=e^{\alpha T},r=Re^{\alpha T},\xi=\phi+\epsilon T,\sigma=\Sigma e^{-\alpha T}\nonumber\\ 
&F=P(R,\xi,v_r,v_\phi;T)e^{-\alpha T},\vec{v}=\vec{Y},\nonumber\\
&\Phi_{dr}=\Phi^{(r)}_{do}\ln{(\alpha R/V)}+\Phi^{(r)}_{do}\delta T+\Psi_{dr}(R,\xi,\theta;T).\label{eq:timedepvars}
\eea
Formally $\alpha$ and $\epsilon$ have the dimension of reciprocal time, but in fact all temporal and spatial quantities (and consequently velocities) may be thought of as numerical values in terms of some fiducial radius $r_o$ and fiducial time $t_o$ when convenient. We have written the potential explicitly in terms of these variables, but by using the various definitions its form is seen to be equivalent to 
\be
\Phi_{dr}\equiv \Phi^{(r)}_{do}\ln{(\alpha r/V)}+\Psi_{dr}(R,\xi,\theta;T).\label{eq:temppot}
\ee
This is the most general form of an isothermally self-similar, non-axially symmetric potential. The coefficient $\Phi^{(r)}_{do}$ is a constant while the function $\Psi_{dr}$ contains the angle, time and radial dependence. The potential corresponds to the unbalanced non-axially symmetric  potential in the locally rotating or `freely-falling' frame. It must satisfy the Laplace equation outside the disc and be consistent with the surface density on the disc. 
  
Through these transformations the independent variables $t,~r,~\phi$ are replaced by the scaled variables $T,~R,~\xi$, while the surface density $\sigma$ is replaced by $\Sigma$ and the scaled DF becomes $P$. The replacement of $\vec{v}$ by unscaled $\vec{Y}$ is peculiar to the isothermality, and is used only to emphasize that the velocity is part of the self-similar scheme. The advantage of using these variables is that, unless the scaled DF $P$ and potential $\Psi_{dr}$ are strictly independent of $T$, the system is general rather than self-similar (\cite{CH91}). This allows us to start with self-similar structure and to follow its evolution as the time dependence arises. This time dependence arises from the winding of the structure.

The CBE must be written entirely in terms of these variables and it becomes successfully 
\bea
\alpha P&=&\partial_TP+(Y_R-\alpha R)\partial_RP+(\epsilon+\frac{Y_\phi}{R}+\frac{V}{\alpha R}\frac{Y_R}{R})\partial_\xi P\nonumber\\
&+&(\frac{Y_\phi^2}{R}+2\frac{V}{R}Y_\phi+\frac{V^2}{R}-\partial_R\Phi_{dr})\partial_{Y_R}P\nonumber\\
&-&(\frac{Y_\phi Y_R}{R}+\frac{VY_R}{R}+\frac{1}{R}\partial_\xi\Phi_{dr})\partial_{Y_\phi}P,
\eea  
  with no loss of generality. The explicit dependence on $T$ must vanish for isothermal self-similarity. 

The solution for $P$ follows from the characteristic equations
\bea
\frac{dP}{dT}&=&\alpha P,\frac{dR}{dT}=Y_R-\alpha R,\nonumber\\
\frac{d\xi}{dT}&=&\epsilon+\frac{Y_\phi}{R}+\left(\frac{V}{\alpha R}\right)\frac{Y_R}{R},\label{eq:timedepchars}\\
\frac{dY_R}{dT}&=&\frac{Y_\phi^2}{R}+\frac{2VY_\phi}{R}+\frac{V^2}{R}-\frac{\Phi^{(r)}_{do}}{R}-\partial_R\Psi_{dr},\nonumber\\
\frac{dY_\phi}{dT}&=&-\frac{1}{R}\left(Y_\phi Y_R+VY_R+\partial_\xi\Psi_{dr}\right),\nonumber
\eea  
 where we have written the potential $\Phi_{dr}$ explicitly.

We observe that if $t,T\leftarrow -t,-T$ and $\alpha\leftarrow -\alpha$, the linear to logarithmic time transformation remains formally unchanged together with the transformations to scaled variables. This is not merely the time-reversed solution at positive time because of the change in sign of $\alpha$. The origin of the solution at $t=0$ is now at $T=+\infty$ and it unfolds in negative time as $T\rightarrow -\infty$. Although the characteristic equations are formally the same, they describe a different development because the velocities and the quantity $\epsilon$ are not changed in sign. They would be changed under a strict time reversal. 

The $Y_\phi$ characteristic equation may be combined with the characteristic expression for $dR/dT$ to give
\be
\frac{d}{dT}\left(\ln{((Y_\phi+V)Re^{\alpha T})}\right)=-\frac{1}{R(V+Y_\phi)}\partial_\xi\Psi_{dr},\label{eq:scaleangmom}
\ee
which in physical variables is the angular momentum equation
\be
\frac{d}{dt}(r(v_\phi+V))=-\partial_\xi\Psi\equiv -\partial_\phi\Phi_{dr}.
\ee
Moreover the $R$, $\xi$ characteristics may be combined with the $Y_R$, $Y_\phi$ characteristics to obtain an energy equation in the form
\bea
\frac{dE'_{dr}}{dT}&=&(\partial_T\Psi_{dr}-\alpha R\partial_R\Psi_{dr}+\epsilon\partial_\xi\Psi_{dr})+\left(\frac{V}{\alpha R}\right)\frac{Y_R}{R}\partial_\xi\Psi_{dr}\label{eq:entemprot}\\
&+&V(V+v_\phi)\left(\frac{d}{dT}(\ln{Re^{\alpha T}})\right).\nonumber
\eea
Here $E'_{dr}\equiv \vec{Y}^2/2+\Phi_{dr}$, where as before $\Phi_{dr}=\Phi^{(r)}_{do}\ln{\alpha R/V}+\alpha \Phi^{(r)}_{do}T+\Psi_{dr}$, is the energy in the locally rotating frame. 

We may also eliminate $Y_R/R\equiv d(\ln{Re^{\alpha T}})/dT$ between this energy equation and equation (\ref{eq:scaleangmom}) to obtain 
\bea
\frac{d(E'_{dr}+V(V+Y_\phi))}{dT}&=&\partial_T\Psi_{dr}-\alpha R\partial_R\Psi_{dr}+\epsilon\partial_\xi\Psi_{dr} \nonumber\\
&+&\!\!\!\left(\frac{V}{\alpha R}\right)(\frac{Y_R}{R}-\alpha)\partial_\xi\Psi_{dr}\label{eq:engtempinert}
\eea
We note that $E_{dr}=E'_{dr}+V(V+Y_\phi)$, which is the energy equal to $\Phi_{dr}+(Y_\phi+V)^2/2+Y_R^2/2$ in the inertial frame but for a constant $-V^2/2$. Thus the last equation can be written as $dE_{dr}/dT$ equal to all the terms on the right that involve $\Psi_{dr}$. 

A swing amplifier  exists between the spiral arm  and the particles, {\it if} the distribution function is a function of $E'_{dr}$. For in such a case the particles are constrained to remain near the arm by the potential $\Phi_{dr}$, and so co-move with the arm on average. We see from equation (\ref{eq:engtempinert}) that we approach this result by requiring all the potential terms on the right hand side to vanish. Once this is applied to equation (\ref{eq:engtempinert}) the inertial energy $E_{dr}$ becomes an exact integral, so that the DF is steady in the inertial frame. In order to have $E'_{dr}$ as an approximate integral in the locally rotating frame, we require in addition from $E_{dr}=E'_{dr}+V(V+Y_\phi)$ that $Y_\phi\ll V$. This is normally the case for the majority of particles in spiral galaxies.

This  condition on the potential 
\be
\partial_T\Psi_{dr}-\alpha R\partial_R\Psi_{dr}+(\epsilon+\frac{V}{\alpha R}(\frac{Y_R}{R}-\alpha))~\partial_\xi\Psi_{dr}=0,\label{eq:potcon}
\ee
is a linear partial differential equation in the potential that imposes a general form . The apparent dependence on $Y_R$ in this equation is only possible for the potential when we remember that the equation holds along a characteristic of the CBE. Hence we may eliminate $Y_R$ using $Y_R=\alpha R+dR/dT$ from the characteristic equations (\ref{eq:timedepchars}). The partial derivatives are evaluated on the CBE characteristic after having been evaluated holding the hidden pair of $(T,\xi,R)$ constant.

The equation so written may be solved by using its own  characteristics to find that $\Psi_{dr}=\Psi_{dr}(r,c)$ where $c\equiv \xi-\epsilon T+V/(\alpha R)$. 
However from our previous definitions of $\xi$, $R$ and $T$ one sees that $c=\phi+Vt/r\equiv \phi+\Omega(r)t$, which is just the inertial azimuthal angle. Thus, as it must be, the potential is steady in the inertial frame where the energy is a constant of the motion.

To obtain a potential that is at least initially compatible with evolving spirals we take it to depend on a spiral combination $\kappa$ of $c$ and $r$ plus $r$ as 
 \be
\Psi_{dr}=\Psi_{dr}(\kappa,r).\label{eq:Psidiscspiral}
\ee
where
\bea
\kappa&=& c+\frac{\epsilon}{\alpha}\ln{r}\equiv \xi+\frac{\epsilon}{\alpha}\ln{R}+\frac{V}{\alpha R},\nonumber\\
&\equiv& \phi+(\frac{\epsilon}{\alpha})\ln{r}+\Omega(r)t,\label{eq:tempspiral}
\eea 
 and note that $\zeta=\Omega(r)t$.

That this is a solution may be confirmed by calculating (using $\kappa(\xi,R)$ as given in equation (\ref{eq:tempspiral}))
\bea
(\partial_T\Psi_{dr})_{R,\xi}&=&\alpha r\partial_r\Psi_{dr},\nonumber\\
-\alpha R\partial_R\Psi_{dr}&=& -\alpha R(-\frac{V}{\alpha R^2}+\frac{\epsilon}{\alpha R})\partial_\kappa\Psi_{dr}-\alpha R\partial_r\Psi_{dr},\\
(\epsilon+\frac{V}{\alpha R^2})\partial_\xi\Psi_{dr}&=&(\epsilon+\frac{V}{\alpha R^2})\partial_\kappa\Psi_{dr}.\nonumber
\eea
The sum of these terms is is $(V/(\alpha R^2)(dR/dT)+\alpha R)\partial_\kappa\Psi_{dr}$, which is zero along along the potential equation characteristic where $dR/dT=-\alpha R$. Along the characteristic of the CBE however $dR/dT=Y_R-\alpha R$, so that to be consistent with the CBE we must regard $Y_R/\alpha R$ as small.  

In fact a more direct approach is simply to neglect $Y_R/(\alpha R)$ compared to one in equation (\ref{eq:potcon}) . Using the definitions of $R$ and $T$, this is equivalent to setting $(Vt/r)(Y_R/V)<1$. As the rotating frame radial velocity is small compared to $V$, this should be well satisfied for sufficiently small $\zeta=Vt/r=V/(\alpha R)$. It transpires that either approach leads ultimately to the same form for $\Psi_{dr}$ with the latter approximation giving $\Psi_{dr}(r,\kappa)$ directly.

In all this discussion we have supposed for brevity that radii are in terms of a fiducial quantity  $r_o=V/\alpha$, and we recall that $\Omega=V/r$. The second independent variable has the form $r=Re^{\delta T}$ in terms of scaled variables. It is through a dependence on this variable that the explicit dependence on $T$ enters, which implies the gradual breaking of the initial spiral self-similarity.  The equivalence $\zeta=\Omega(r)t=V/(\alpha R)$ is helpful when interpreting the behaviour found below. 

With this form of the potential and the particle velocities in the locally rotating frame being small, equation (\ref{eq:engtempinert}) gives $E'_{dr}$ as an approximate integral of the CBE in the  locally rotating frame.  Since $E'_{dr}$ is an approximate integral we expect from the first of the CBE characteristics  (\ref{eq:timedepchars}) that $P=F(E'_{dr})e^{\alpha T}$. But $\Sigma=\int~P~dY_RdY_\phi$ and this should be independent of $T$ for initial self-similarity. Thus, recalling the form of the potential in scaled variables as $\Phi_{dr}=\Phi^{(r)}_{do}\ln(\alpha R/V)+\alpha\Phi^{(r)}_{do}T+\Psi_{dr}$ plus $E'_{dr}=\vec{Y}^2/2+\Phi_{dr}$,  we see that an isothermal DF in the locally rotating frame such as 
\be
F(E'_{dr})=K_{dr}\exp{\left(-\frac{E'_{dr}}{\Phi^{(r)}_{do}}\right)},\label{eq:DFtransspiral}
\ee
removes the $T$ dependence until it develops in the asymmetric potential $\Psi_{dr}$.

Here $K_{dr}$ is the normalization for the transient, asymmetric, collisionless distribution function.  The mean velocity of these particles is zero in the locally rotating frame, due to the symmetry of the isothermal DF. For the approximation $|\vec{Y}|<V$ to hold we require 
\be
\Phi^{(r)}_{do}\le V^2.\label{eq:confincon}
\ee

 To the extent that this rather weak condition holds, the isothermal DF should continue to be a solution of the CBE in the potential given by equation (\ref{eq:rotPoiss}) below. 
By treating the DF of the particles we avoid detailed consideration of their orbits, but the characteristic equations (\ref{eq:timedepchars}) do provide these if necessary.

The potential form (\ref{eq:Psidiscspiral}) must be made compatible with the Laplace equation above the disc.  
namely 
 \bea
\!\!\!\!\!\!\!\!\!\!\!\!\!\!\!\!\!&~&\frac{1}{r^2}(\partial_r(r^2\partial_r(\Phi))\nonumber\\
\!\!\!\!\!\!\!\!\!\!\!\!\!\!\!&+&\frac{1}{\sin^2{\theta}}\partial_\theta(\sin{\theta}\partial_\theta\Phi)+\frac{1}{\sin^2{\theta}}\partial^2_\phi~\Phi)=0.
\eea
We do this by assuming a $\theta$ dependence ($\theta$ is another scaled variable since it is dimensionless) in $\Psi_{dr}$. 
After inserting the alternate form for the asymmetric potential (the axi-symmetric disc potential already satisfies the Laplace equation above the disc) namely
\be
\Phi_{dr}=\Phi^{(r)}_{do}\ln{r}+\Psi_{dr}(\kappa,\theta,r),
\ee
the Laplace equation becomes the master equation for our purposes in this paper 
\bea
&~&\Phi^{(r)}_{do}+\frac{\epsilon}{\alpha}\partial_\kappa\Psi_{dr}+\partial_r(r^2\partial_r\Psi_{dr})\nonumber\\
&+&(\frac{\epsilon}{\alpha}-\frac{Vt}{r})r\partial_r\partial_\kappa\Psi_{dr}\nonumber\\
&+&\left((\frac{\epsilon}{\alpha}-\frac{Vt}{r})^2+\frac{1}{\sin^2{\theta}}\right)\partial^2_\kappa\Psi_{dr}\nonumber\\
&+&\frac{1}{\sin{\theta}}\partial_\theta(\sin{\theta}\partial_\theta\Psi_{dr})=0.\label{eq:rotPoiss}
\eea

In the Laplace equation the variables $t$ and $r$ are independent. Hence we can replace $r$ everywhere by $\zeta=Vt/r$ since $r\partial_r=-\zeta\partial_\zeta$ and $\partial_r(r^r\partial_r)=\zeta^2\partial_\zeta^2$. we will use this form in finding the solution for $\Psi_{dr}$. We can also note in passing from this equation that one can  neglect the $r$ dependence in $\Psi_{dr}$ at any given $t$, and so preserve strict initial spiral symmetry ($\kappa$ dependence only), {\it only} if $\zeta=Vt/r\equiv \Omega(r)t<\epsilon/\delta$.  This is generally as expected, but the dependence on the tangent of the initial winding angle $\epsilon/\delta$ is of interest.   

The potential equation (\ref{eq:rotPoiss}) is linear and can be solved in terms of base functions in the form
\be
\Psi^m_{dr}=\Phi^{(r)}_{do}\ln{\sin{\theta}}+e^{(im\kappa)}T_m(\theta){\cal R}_m(\zeta).\label{eq:Psid}
\ee
We recall that the phase $\kappa=\phi+(\epsilon/\delta)\ln{r}+Vt/r$ and $\phi$ is in the locally rotating frame. Naively, by ignoring the $\zeta$ dependence in the potential, the log spiral is completely wound up at a fixed $r$ when $\Omega(r) t=2\pi$. This gives $\approx 10^{7.5}$ years at $r=10$ kpc and $V=200$ km/sec. However even in this naive limit there is an outward moving `winding wave', given by $\zeta\equiv Vt/r=cst\ll(\epsilon/\delta)\ln{r}$, outside of which the pure log spiral remains recognizable and similarity is maintained. But as time progresses the $\zeta$ dependence can not be ignored and it is the consequence of this evolution away from the initial self-similar spiral arm that will occupy us in this paper.

The  axi-symmetric background should include both a Mestel disc and an isothermal halo. The complete potential for such structure is well known (e.g. \cite{BK75}, \cite{MRS81}, \cite{T82}), and a self-similar derivation is given in (\cite{Harxiv2012}). In general the halo also possesses a non axi-symmetric component. However  we ignore  the non axi-symmetric halo here as it is likely to be a small effect at the disc. 

The axi-symmetric isothermal halo potential at the disc is of the form $\Phi_{dh}=\Phi^{(h)}_{do}\ln{r}$ with $\Phi^{(h)}_{do}$ constant, but it is  removed in the locally rotating frame of the disc. The coefficient $\Phi^{(r)}_{do}$ gives the radial potential component of the initial self-similar spiral structure. We will find that it should be substantially  larger than the non-axially symmetric potential $\Psi_{dr}$.   

The Mestel Disc potential takes the form (e.g. \cite{Harxiv2012})
\be
\Phi_a=\frac{2\pi G \Sigma_a}{\delta}\left(\ln{\delta r}+\ln{(1+\cos{\theta})}\right),\label{eq:Mpot}
\ee
where we have set $\delta\equiv \alpha/V$ and the axi-symmetric surface density is $\sigma_a=\Sigma_a/(\delta r)$. In the plane the total inertial potential becomes 
\be
\Phi=\Phi_a(\pi/2)+\Phi_{dr}+\Phi_{dh}=\Phi_{oa}\ln{\delta r}+\Psi_{dr}\equiv (\frac{2\pi G \Sigma_a}{\delta}+\Phi^{(r)}_{do}+\Phi^{(h)}_{do})\ln{\delta r}+\Psi_{dr}.\label{eq:Mplapot}
\ee
 
In the local disc reference frame however, we may focus on the non-axi-symmetric spiral potential (including the axi-symmetric part) since 
\be
V^2=\frac{2\pi G \Sigma_a}{\delta}+\Phi^{(h)}_{do}.\label{eq:freefall}
\ee  
This free-fall condition removes the influence of the undisturbed disc-halo potential on the spiral particles. The developing spiral structure perturbs the radial and angular structure of the disc through the action of $\Phi_{dr}$.

The distribution function of the axi-symmetric particles is not our concern in this paper but it is also approximated by an isothermal distribution in the rotating frame (\cite{Harxiv2012}). The total axi-symmetric surface density may also contain isothermal gas, for which a collisionless DF is not relevant.

\section{Potential-Disc Coupling}

The asymmetric potential found from the Laplace equation (\ref{eq:rotPoiss}) must obey the disc boundary condition (the axi-symmetric part is cancelled by $-(1/r)\partial_\theta\Phi_a|_{\theta=\pi/2}$) in the absence of gas as
\be
2\pi G\sigma_{dr}=-\frac{1}{r}(\partial_\theta\Psi_{dr})|_{\theta=\pi/2},\label{eq:discsigma}
\ee   
where 
\be
\sigma_{dr}=\int~F(E'_d)~dY_RdY_\phi\equiv 2\pi \Phi^{(r)}_{do} K_{dr}\frac{e^{-\frac{\Psi_{dr}}{\Phi^{(r)}_{do}}}}{r}.\label{eq:collsigmapot}
\ee
This is equivalent to treating the disc as a volume density with a delta function  $\delta(z)$ or $\delta(\cos{\theta})/r$ in the Poisson equation. Integrating about $z=0$ or $\theta=\pi/2$ gives this boundary condition on the upper half space.
 
This condition is generally difficult to satisfy for all $\kappa$ for a single  value of $m$, because of the exponential dependence of $\sigma$ on the potential that follows from the last integral. Fortunately we can arrange to satisfy it everywhere by adding isothermal gas on the left of equation (\ref{eq:discsigma}) to obtain the augmented equation (\ref{eq:gas+disc}) below. The gas is not described by the isothermal collisionless DF and so provides an additional degree of freedom. In principle its distribution is determined by the magnetohydrodynamic equations in the potential $\Psi_{dr}$, but we avoid this in this paper by simply using the augmented equation (\ref{eq:discsigma}) as an equation for the gas surface density.

Equation (\ref{eq:rotPoiss}) with the base ans\"atz of equation (\ref{eq:Psid}) is resolved into two equations ($\epsilon\leftarrow \epsilon/\alpha$ for compactness)
\bea
&~&\!\!\!\!\!\!\!\!\!\!\!\!\!\frac{1}{\sin{\theta}}\frac{d}{d\theta}(\sin{\theta}\frac{dT_m}{d\theta})+T_m(\theta)(k_m^2-m^2(\epsilon^2+\frac{1}{\sin^2{\theta}})+im\epsilon)\nonumber\\
&=&0,\label{eq:sep1}\\
&~&\!\!\!\!\!\!\!\!\!\!\!\!\!\zeta^2\frac{d^2{\cal R}_m}{d\zeta^2}-im\zeta(\epsilon-\zeta)\frac{d{\cal R}_m}{d\zeta}-{\cal R}_m(m^2(\epsilon-\zeta)^2+k_m^2-m^2\epsilon^2)\nonumber\\
&=&0,\label{eq:sep2}
\eea  
where $k_m^2$ is the separation constant (positive or negative or zero) and $\zeta\equiv Vt/r$.

The base solution for the radial dependence takes the form  
\bea
{\cal R}_m(\zeta)&=&\exp{i(-\frac{m\zeta}{2}+\frac{m\epsilon}{2}\ln{\zeta})}(A_{1m}M_{\lambda,\mu}(\sqrt{3}m\zeta)\nonumber\\
&+&A_{2m}W_{\lambda,\mu}(\sqrt{3}m\zeta)),\label{eq:asymrad1}
\eea
where $M,W$ are Whittaker functions, $\lambda\equiv \sqrt{3}m\epsilon/2$, and $\mu\equiv (\sqrt{(1+im\epsilon)^2+4k_m^2})/2$. 
One must remember that this asymmetric solution is to be added to an axially symmetric Mestel disc or disc-halo (e.g. \cite{Harxiv2012}), so that positive and negative values are acceptable. The amplitude must however be less than that of the axially symmetric potential.

\begin{figure}[]
\begin{tabular}{cc} 
{\rotatebox{0}{\scalebox{.4}
{\includegraphics{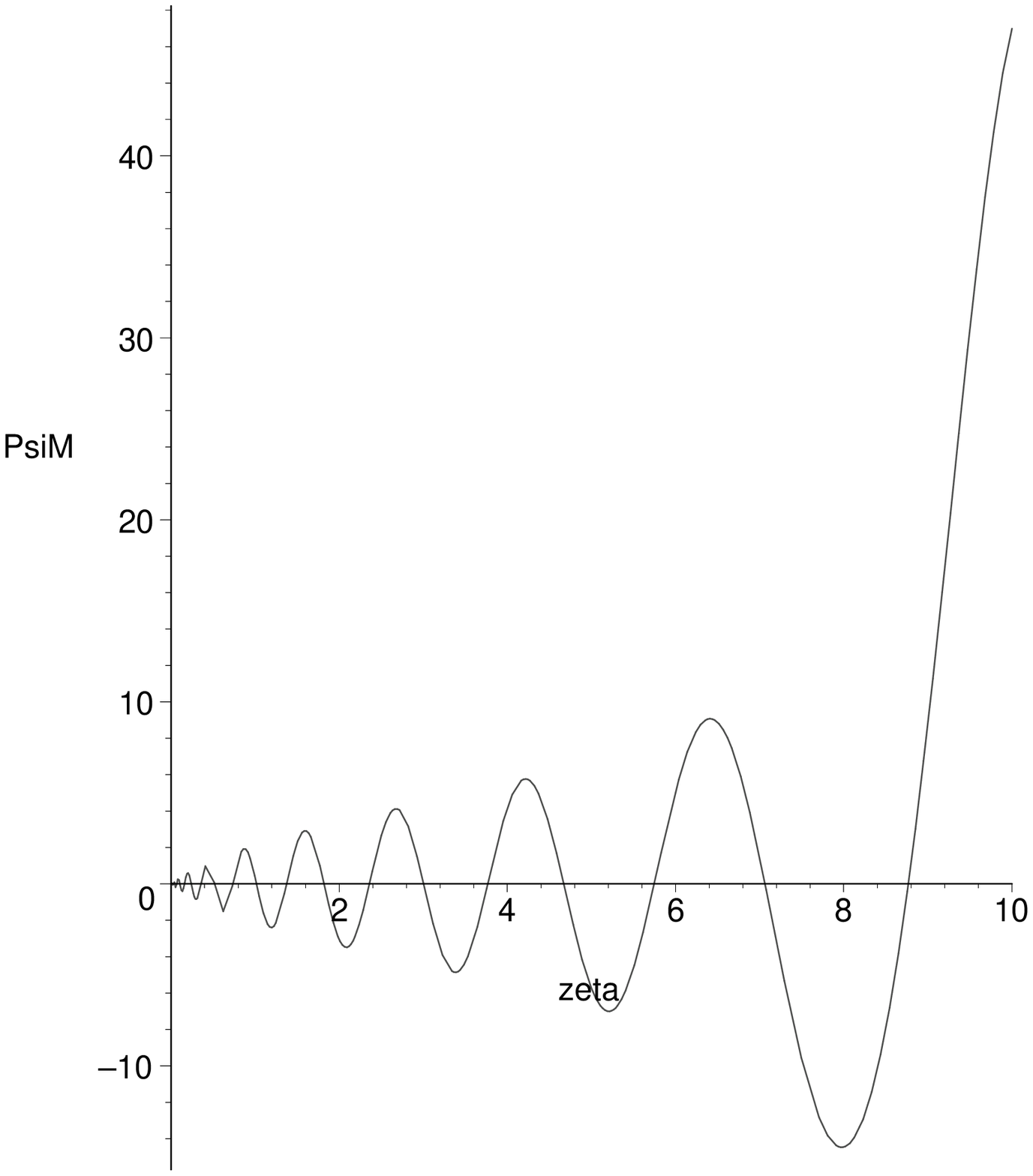}}}}&
{\rotatebox{0}{\scalebox{.4}
{\includegraphics{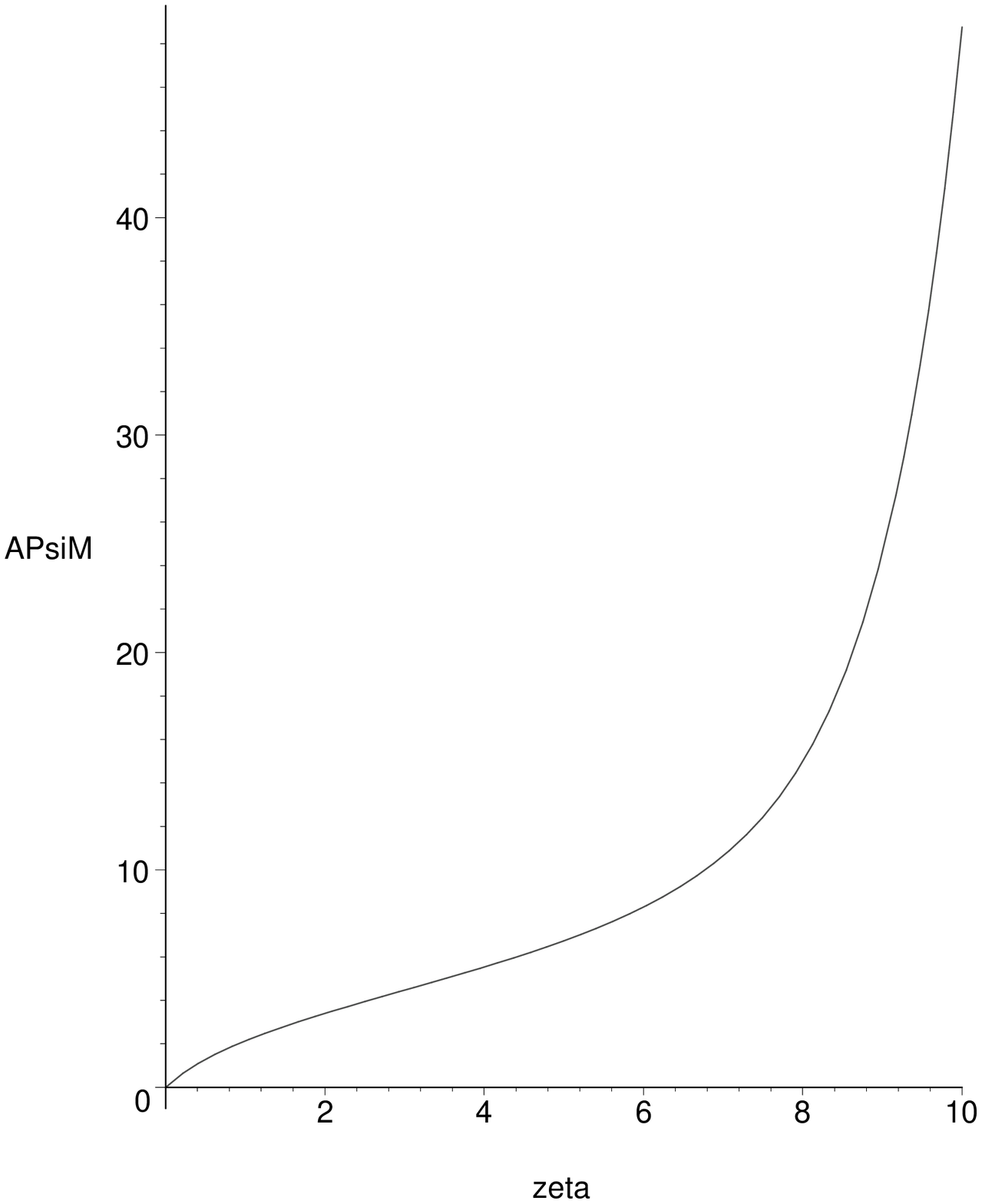}}}}\\
{\rotatebox{0}{\scalebox{.4} 
{\includegraphics{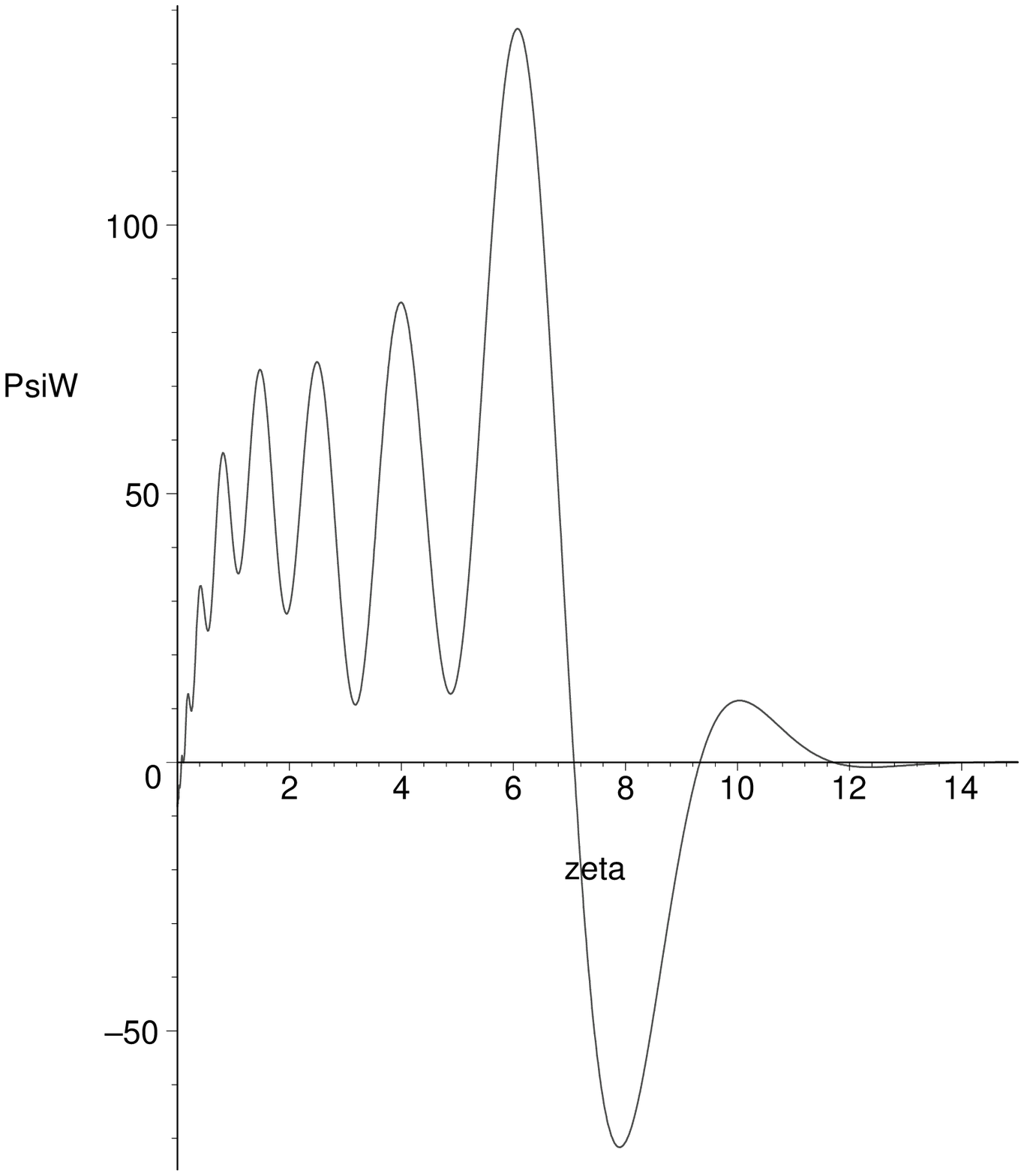}}}}&
{\rotatebox{0}{\scalebox{.4}
{\includegraphics{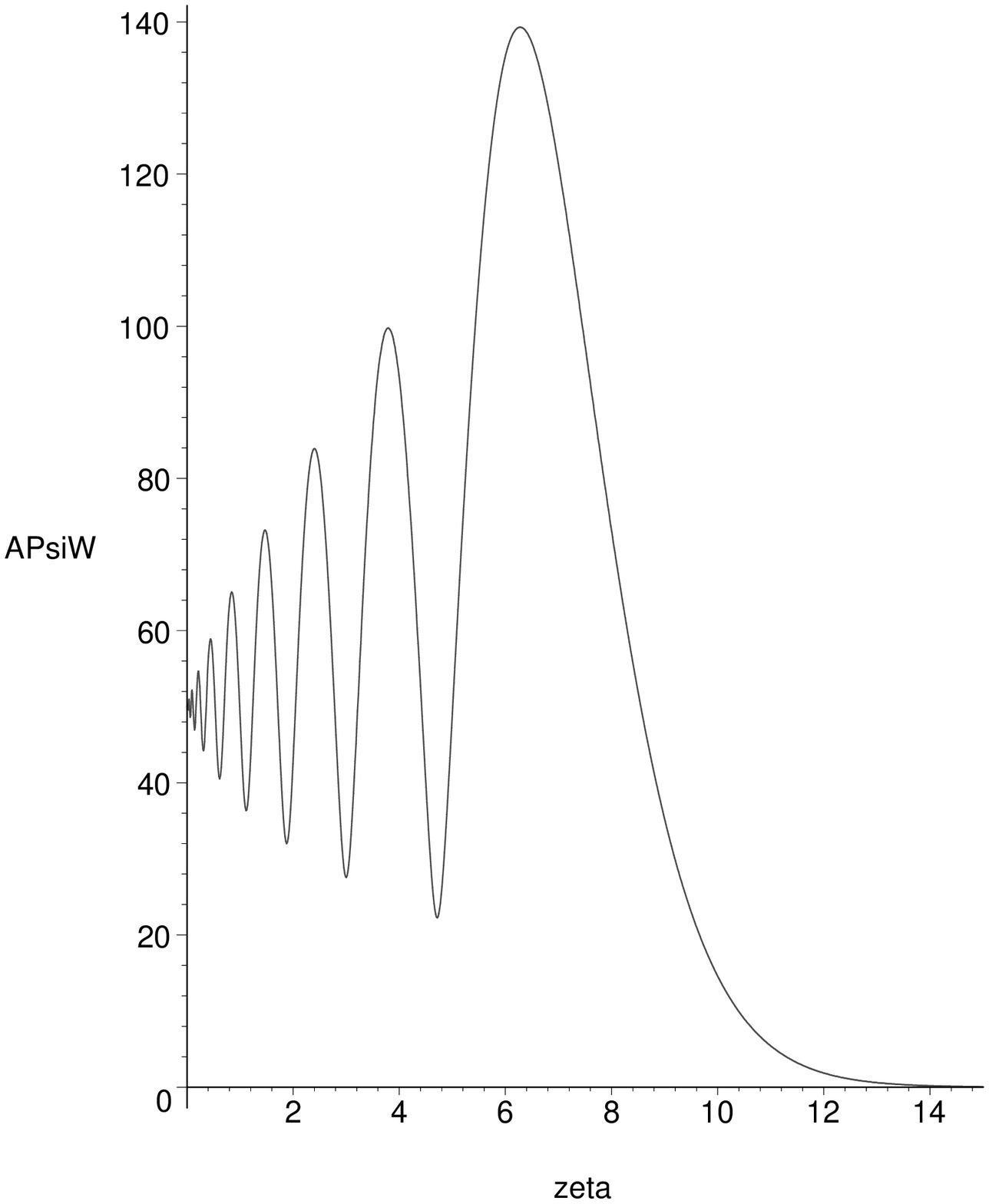}}}}\\
\end{tabular}
\caption{On the upper left we show the real part of the asymmetric potential proportional to the whittaker M function, as a function of $\zeta$ for $\nu=0$, $m=2$, and $\epsilon/\alpha=3.5$. The amplitude constant $T_2(\pi/2)A_{12}$  is omitted. The upper right curve is the modulus of this potential component. At lower left we show the real part of the potential proportional to the Whittaker W function for the same parameters ($T_2(\pi/2)A_{22}$ is  omitted). At lower right is the amplitude of this potential component. The same graph as at upper left for a leading spiral with $\epsilon/\alpha=-3.5$ diverges much more rapidly starting at $\zeta\approx 0.7$.  In every case $k_m=0$.}     
\label{fig:potcomps}
\end{figure}

Our base solutions are not normal modes since they are not standing waves. However in the full Mestel disc without imposed boundaries they are the modes available to the disc.
The various possible asymmetric potential base behaviours as a function of $\zeta$ can be seen for typical values in figure(\ref{fig:potcomps}). We have applied the exponential phase factor $e^{im\kappa}$ in equation (\ref{eq:Psid}) to the potential. One should recall (from equation (\ref{eq:tempspiral})) that we can write $\kappa=\nu+\zeta$, where $\nu\equiv \phi+(\epsilon/\alpha)\ln{r}$ is the undisturbed logarithmic spiral for $\nu$ constant and $\zeta\equiv \Omega t$. We show only the real part and amplitude of the functions, but the imaginary part might also be relevant if the multiplicative factors are imaginary. 

 We see from the figure  that the `M term' represents an oscillation growing in $\zeta$ from zero amplitude. It is ultimately unstable and grows exponentially after $\zeta\approx 8$. The case that we illustrate has $m=2$ and $\epsilon/\alpha=3.5$. A similar calculation with $\epsilon/\alpha=2.5$ reveals that this instability is similarly violent. It becomes more violent with significantly decreasing winding angle. 

However it is clear that for any $t>0$ this term is singular at the origin since $\zeta$ will be infinite there. This singularity is  evidently the formal source of the instability since the value at any fixed $\zeta$ propagates to larger radius in time. However we must remember that we require $\zeta Y_R/V\ll1$ for our potential solution to be valid. Thus there is only a limited range of $\zeta$ accessible to our solutions and infinite $\zeta$ is  excluded unless $Y_R$ tends to zero at small radius.  Generally there is an effective inner boundary at finite $t$, set by some maximum $\zeta$ that depends on the maximum value of $Y_R/V$. This also limits the growth in time to a maximum $\zeta$ at a given radius unless $Y_R=0$ everywhere.   

 The potential term proportional to the `W function' represents the development from an initial spiral of fixed $\nu$ of finite amplitude. The growth is also oscillatory (in both amplitude and real value) but rather than growing exponentially, it is exponentially damped after passing through a strong localized amplification, which we identify with swing amplification. In contrast to the M term, we find that the amplification becomes rapidly weaker as the winding angle $\epsilon/\alpha$ declines.  

 In this paper we focus on the W function behaviour since the origin of the initial arm may be left undetermined. The history that we study is then a transient swing amplification  that ends in an exponential decay. The potential determines the arm surface density through equation (\ref{eq:discsigma}).

We choose the  separation constant to be $k_m=0$ as an initial condition that yields an initial spiral in the Whittaker W base mode. This is so because, provided that $Re(2\mu+1)\ge 2$ and uniquely when $k_m=0$, the term in the potential proportional to the W function goes to (e.g. \cite{AS72}) 
\be
\left(A_{2m}T_m(\pi/2)e^{-im\epsilon\ln{(\sqrt3)}/2}\frac{\Gamma(2\mu)}{\Gamma(\mu-\lambda+1/2)})+cst\times z\times e^{im\epsilon\ln{(z)}/2}\right)e^{im\nu},
\ee
as $\zeta\rightarrow 0$. Here the argument $\sqrt{3}m\zeta\equiv z$. The product of $A_{2m}T_m(\pi/2)\exp{(-im\epsilon\ln{(\sqrt{3})}/2)}$ with the ratio of the Gamma functions is a complex constant, wherein $\lambda$ is $\sqrt{3}m\epsilon/2$ as above, and now $\mu=(1+im\epsilon)/2$ since $k_m=0$. This yields our  initial condition of a pre-existing arm at $\zeta=0$ since the latter expression for the limiting W potential tends to this complex constant as $\zeta\rightarrow 0$.

When $k_m=0$ the solution for $T_m(\theta)$ is given in terms of the familiar associated Legendre functions as
\be
T_m(\theta)=(C_{1m}~P^m_{im\epsilon}(x)+C_{2m}~Q^m_{im\epsilon}(x)),\label{eq:initspiral}
\ee  
where $P^\mu_\nu$ and $Q^\mu_\nu$ denote the associated Legendre functions and $C_{1m}$, $C_{2m}$ are  complex constants.

The behaviour that interests us most is that for $\zeta\ge 1$,  when the evolution of the initial spiral is marked. Using the asymptotic forms applicable when $z=\sqrt{3}m\zeta>1$ \cite{AS72} the potential at the disc in this regime takes the form (we can absorb $A_{2m}$ into the constants in $T_m$)
\be
\Psi_{dr}\simeq T_m(\pi/2)\exp{im(\nu+\zeta/2+\epsilon\ln{\zeta}/2)}z^{\lambda}e^{-z/2}.\label{eq:evolvpot}
\ee
We see that the phase of the original spiral $\nu$ is modified by the extra terms depending on $\zeta$ in the exponential plus any phase constant in $T_m(\pi/2)$. The amplitude is modulated in $\zeta$ by the function $z^\lambda e^{-z/2}$, which peaks at $z=2\lambda$ or $\zeta=\epsilon/\alpha$. \footnote{We recall that $\epsilon$ in the formulae denotes the initial spiral winding angle $\epsilon/\alpha$.} This last value should be $\ge O(1)$ for the peak to lie in the asymptotic range. The value at this peak is $e^{\lambda(\ln{2\lambda}-1)}$, which can be extremely large. This is the growth that we identify with a non-linear version of the `swing amplification' (\cite{T82}). 

The `swinging' in the evolution can be seen by setting the change in the total phase of the potential to zero. This identifies the original spiral in the asymptotic region where it is evolving in time and space. Using the phase in equation (\ref{eq:evolvpot}) we find from the zero differential 
\be
d\nu=-\frac{d\zeta}{2}(1+\frac{\epsilon}{\alpha}\frac{1}{\zeta}).\label{eq:dnu}
\ee
We recall that $d\nu=d\phi+(\epsilon/\alpha) dr/r$, and we have restored $\epsilon$ to $\epsilon/\alpha$. At fixed $r$, $d\nu=d\phi$ and $d\zeta=Vdt/r$, so that the previous equation implies
\be
r\frac{d\phi}{dt}=-\frac{V}{2}(1+\frac{\epsilon}{\alpha}\frac{1}{\zeta}).\label{eq:dphidt}
\ee
Hence an initially trailing arm ($\epsilon/\alpha>0$) is winding up in the trailing sense at each $r$ with an approximate maximum rate at $\zeta=\epsilon/\alpha$ of $-V$. An initially leading arm ($\epsilon/\alpha<0$) swings from a winding forward rate to a winding backwards rate at the  peak amplification where $\zeta=|\epsilon/\alpha|$, and continues at large $\zeta$ to reach the limiting trailing rate of $-V/2$ thereafter.

The true shape of the spiral at any instant is extended in $r$ at fixed $t$. From that perspective, $d\zeta=-(\zeta/r)dr$ and $d\nu=d\phi+(\epsilon/\alpha)(dr/r)$, and hence equation (\ref{eq:dnu}) yields
\be
r\frac{d\phi}{dr}=\frac{1}{2}(\zeta-\frac{\epsilon}{\alpha}).\label{eq:dphidr}
\ee
This implies that an initially trailing wave ($\epsilon/\alpha>0$) is now leading at small $r$ (large $\zeta$) but still trails at large $r$ (small $\zeta$). The transition is again at the peak value $\Omega t\equiv \zeta=\epsilon/\alpha$ or at $t=r(\epsilon/\alpha)/V$. An initially leading spiral ($\epsilon/\alpha <0$) is tightly wound at small $r$ (large $\zeta$) and is more open  at large $r$ (small $\zeta$).  We can not extrapolate strictly to very small $\zeta$ since equation (\ref{eq:evolvpot}) holds for large $z=\sqrt{3}m\zeta$, but either side of the peak value is permitted.    

\begin{figure}
\begin{tabular}{cc} 
{\rotatebox{0}{\scalebox{.4}
{\includegraphics{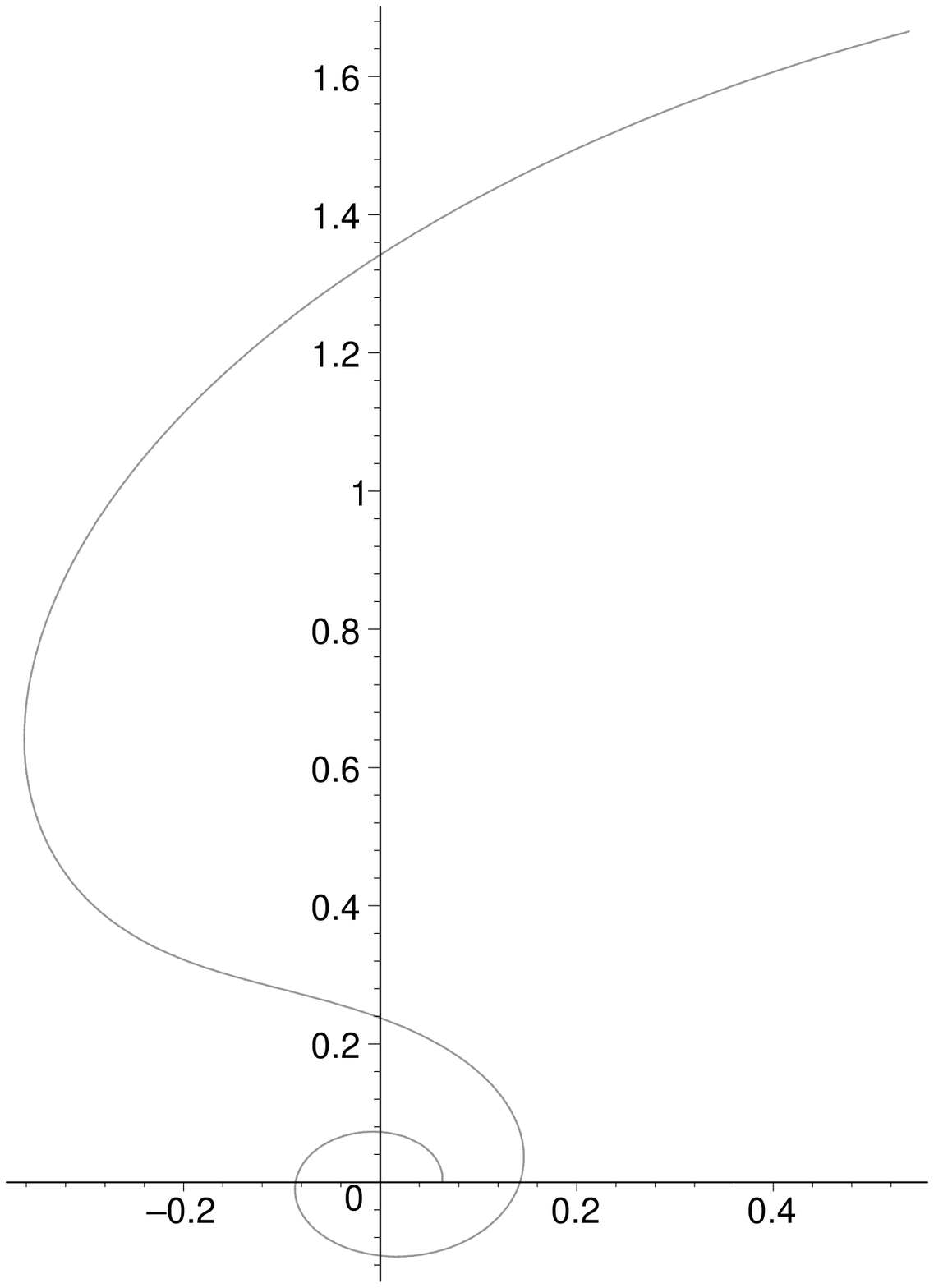}}}}&
{\rotatebox{0}{\scalebox{.4}
{\includegraphics{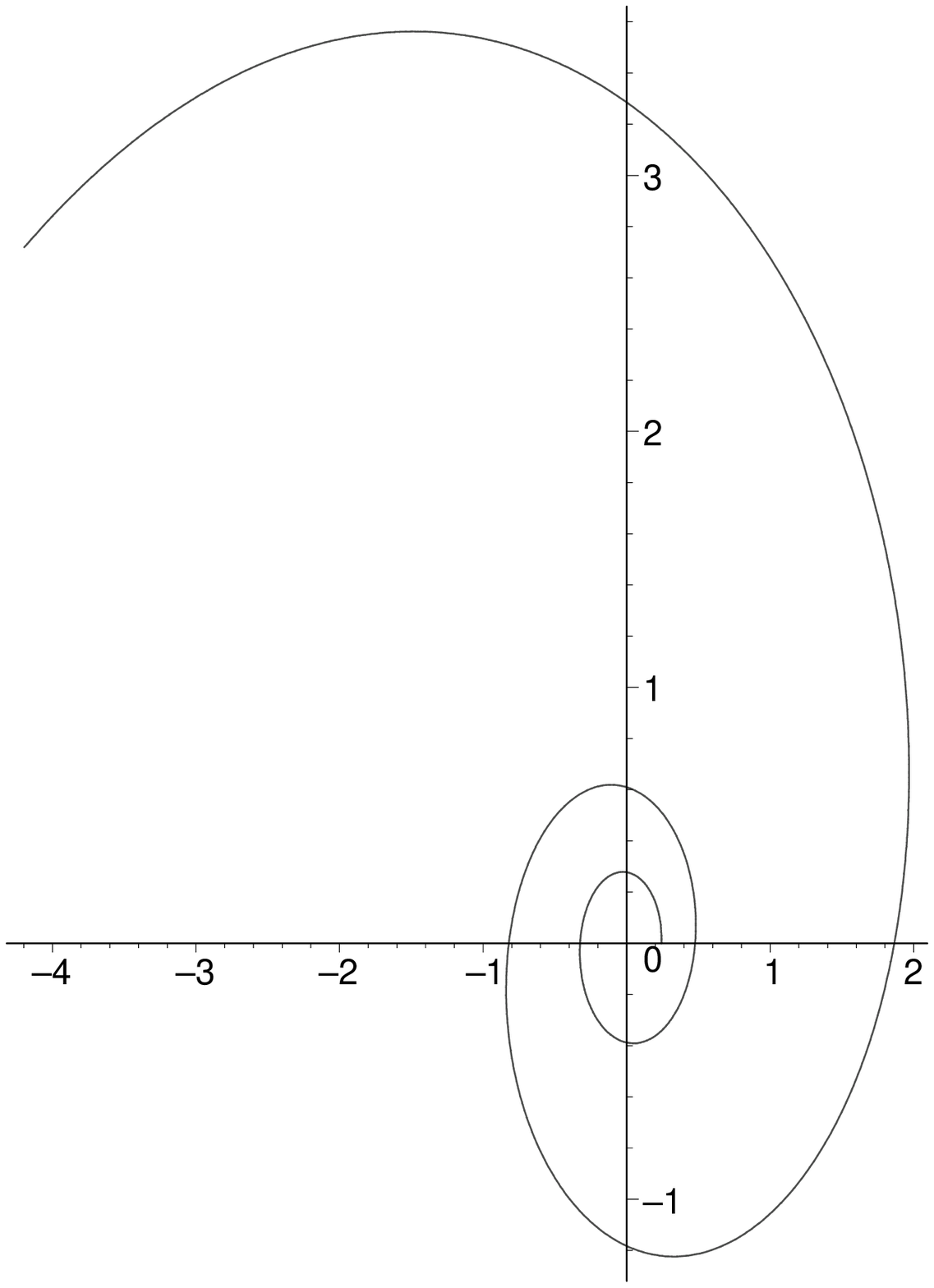}}}}\\
{\rotatebox{0}{\scalebox{.4} 
{\includegraphics{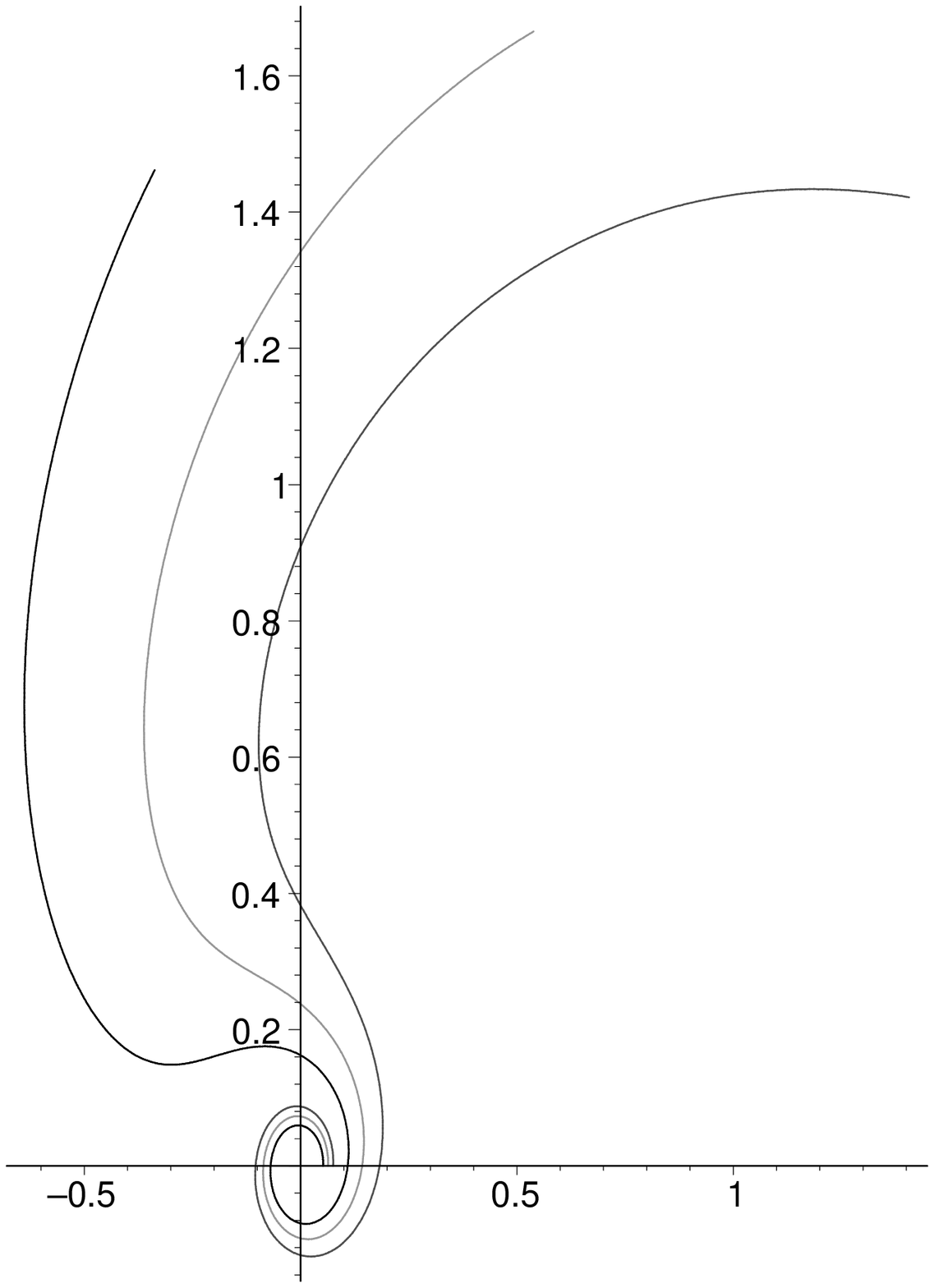}}}}&
{\rotatebox{0}{\scalebox{.4}
{\includegraphics{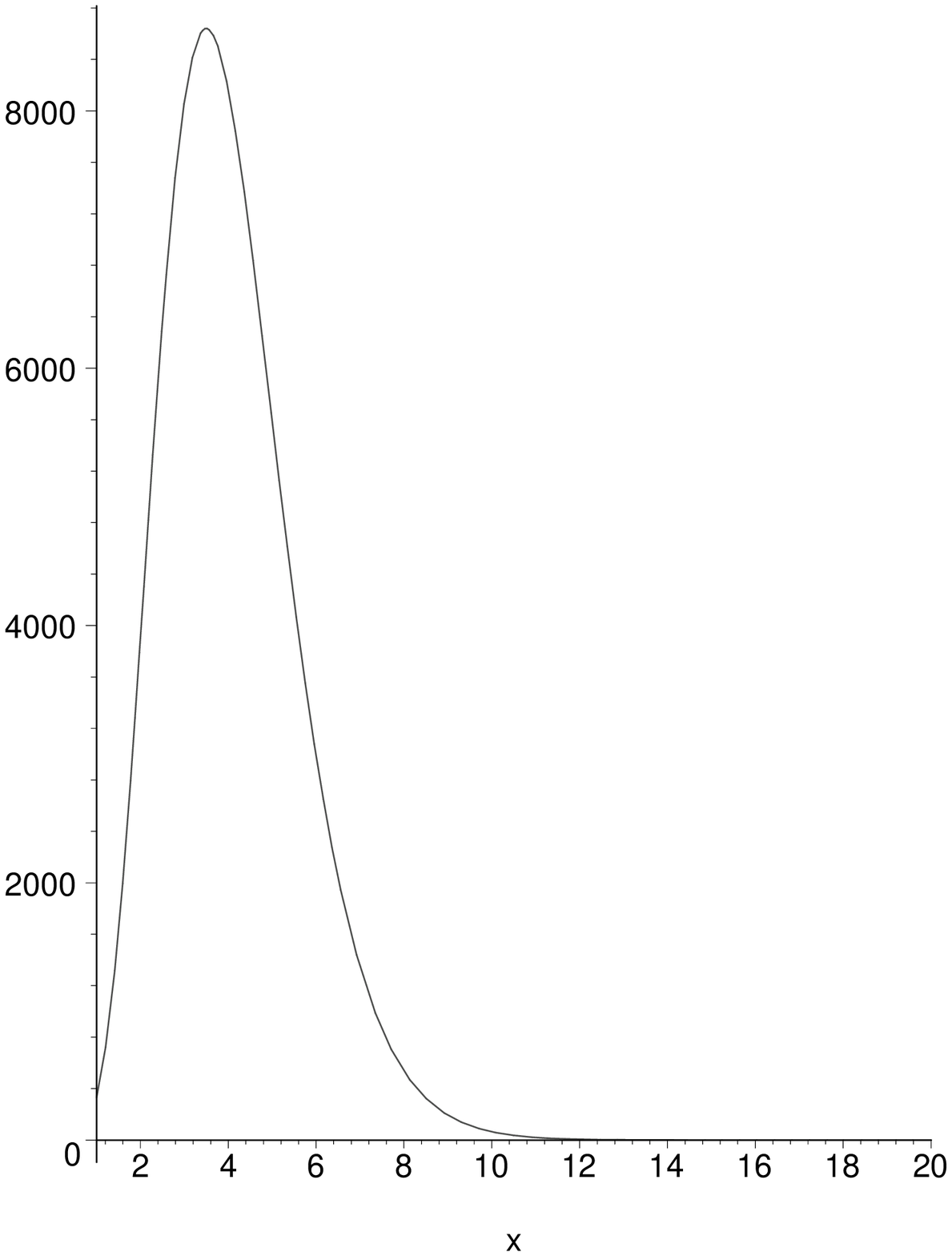}}}}\\
\end{tabular}
\caption{The figure on the upper left shows a spiral with initial winding angle $\epsilon/\alpha=3.5$ and $Vt=1.75$ in the $xy$ plane. The arbitrary numerical value of the total phase is set equal to $10$ for the illustration.  The figure on the right shows a leading spiral with the same winding angle  at $Vt=5$. The bottom left figure shows the same trailing spiral at the sequence of times $Vt=1.5,~1.75,~2.0$. The spiral moves out and swings back as time advances. The bottom right figure shows the amplitude modulation function from equation (\ref{eq:evolvpot}) for $m=2$ and $\epsilon=3.5$ as a function of $\zeta$.}    
\label{fig:windingspirals}
\end{figure}

This behaviour is indicated in figure (\ref{fig:windingspirals}), where the phase from equation (\ref{eq:evolvpot}) is plotted. We see there that at small $r$ we have indeed a leading spiral, which transits to a trailing spiral at $\zeta=\epsilon/\alpha$, the swinging amplifier peak. An initially leading spiral unwinds slowly from the centre outwards as expected.

The independent self-similar evolution with $\alpha<0$ and $\zeta=-V|t|/r$ is complementary to the solution above. Considering the structure in radius, we see from equation (\ref{eq:dphidr}) that an initially trailing spiral ($\epsilon/\alpha>0)$ trails everywhere, but most tightly at small radius. An initially leading spiral ($\epsilon/\alpha<0$) is trailing at small $r$ and leading at large $r$, with the transition occurring again at $\zeta=|\epsilon/\alpha|$. Thus the behaviours of the leading and trailing spirals are interchanged from the case with $\alpha>0$. This is also the case with the rates from equation (\ref{eq:dphidt}). The interchange of leading and trailing spirals with different sign of the winding angle $\epsilon/\alpha$ at small $r$, suggests the possibility of the spiral structures arising from waves passing through the centre of the system (e.g. \cite{BT08},pp 512-513).

Such behaviour might correspond to the self-excited M term in the potential, if the spiral waves are reflected at spatial infinity where $\zeta=0$ and so the amplitude is zero there. It still requires $Y_R\rightarrow 0$ for the centre to lie within the range of our approximation. Moreover the swing amplification must operate on the leading and trailing waves. In fact the amplification seems to much more rapid for leading waves (see the caption of figure (\ref{fig:potcomps})).

\begin{figure}[t]
\begin{tabular}{cc} 
{\rotatebox{0}{\scalebox{.4}
{\includegraphics{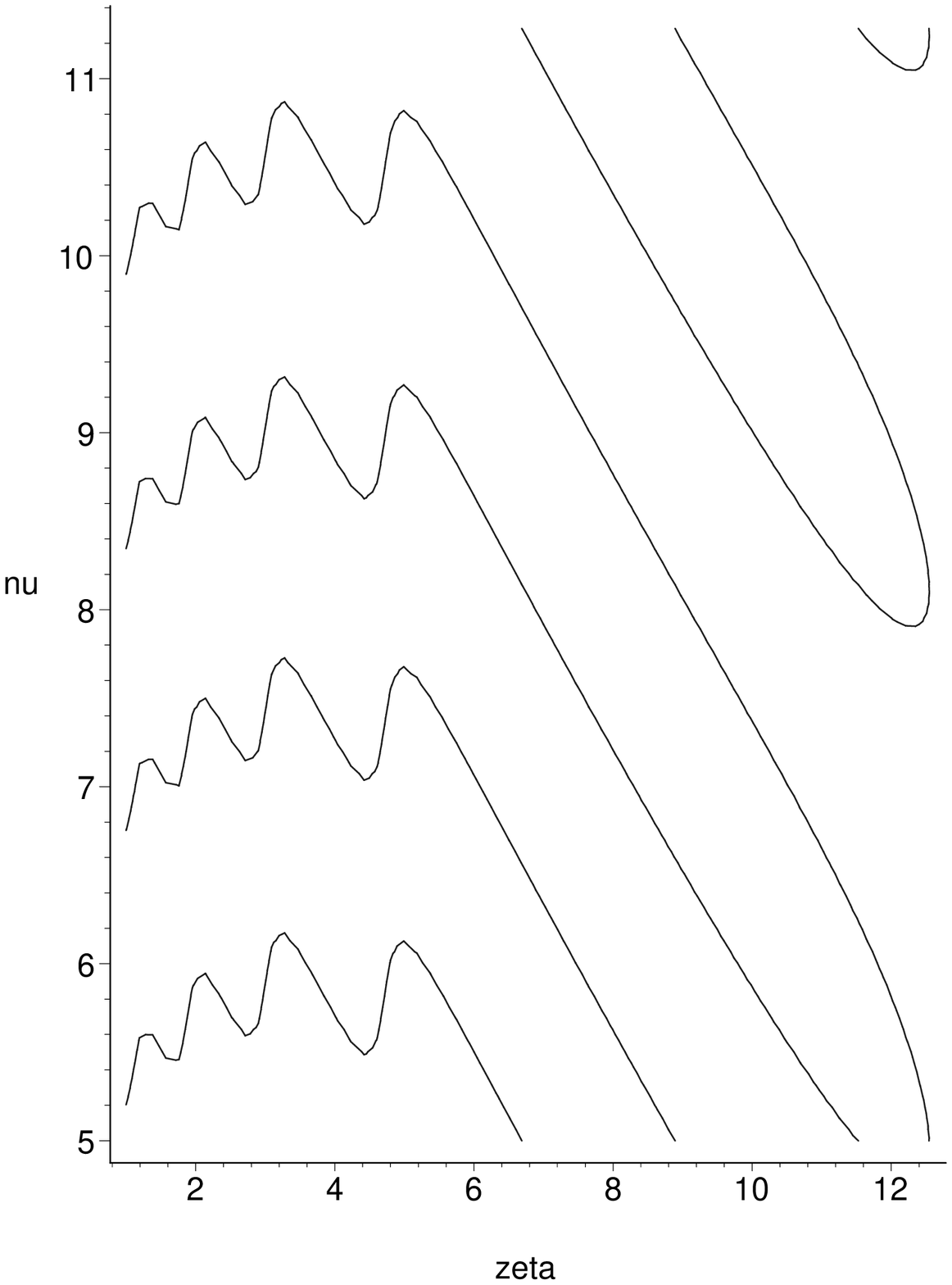}}}}&
{\rotatebox{0}{\scalebox{.4}
{\includegraphics{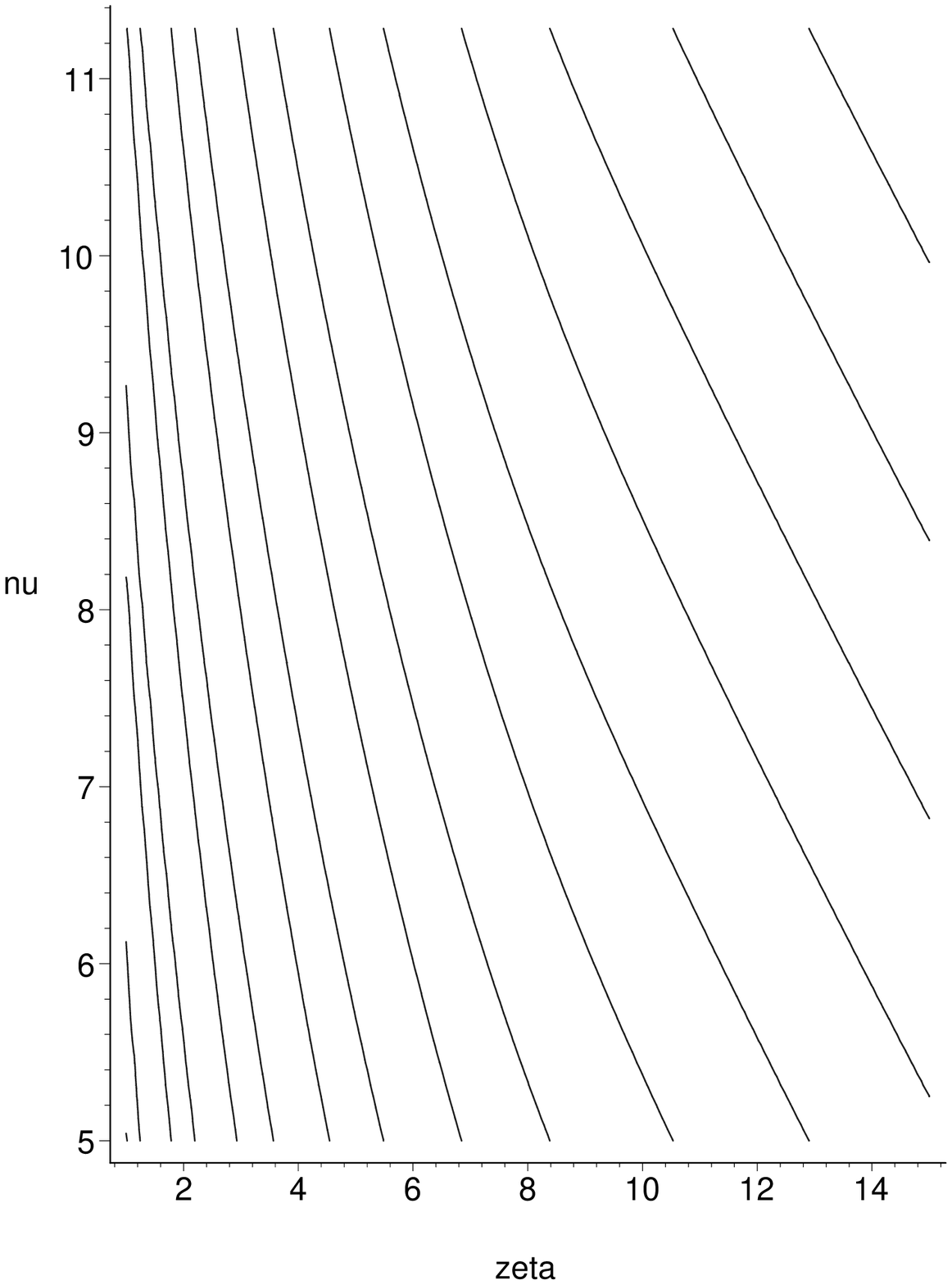}}}}\\
\end{tabular}
\caption{The figure  shows on the left $\nu$ as a function of $\zeta$ when the real part of the potential is $1$ for the W component. On the right we show the curves when the real part of the potential is set equal to $1$ for the M component.  We set $m=2$ and $\epsilon/\alpha=3.5$ in all cases.  }    
\label{fig:nuzeta}
\end{figure}

We note that the inner spirals, either leading or trailing are distorted to the opposite winding at a radius $r=Vt/(\epsilon/\alpha)$. This radius is $10$ kpc after $0.18$ Gyr if the winding angle $\epsilon/\alpha=3.5$. The initial pitch angle is about $16^\circ$. This interval is just over one half of a rotation period at this radius, but the outer spiral is not destroyed by this time.
The oscillating nature of the potential seen in figure (\ref{fig:nuzeta}) below  corresponds to the winding of the arm in radius and the outwards propagation of the arm in time. When $m=2$, any value of $\nu$ plus $\nu+\pi$ represent the two original arms.  

The bottom right curve in figure (\ref{fig:windingspirals}) shows the swing amplification amplitude when $\epsilon/\alpha=3.5$, and $m=2$, as a function of $\zeta$. The maximum is pronounced in this case, but it declines sharply with declining $\epsilon/\alpha$ and $m$. Thus when $\epsilon/\alpha=1$ (so that the pitch angle and the winding angle are equal at $45^\circ$), the maximum is $\approx 1.52$. At $\epsilon/\alpha=2$ (winding angle $63.4^\circ$ it is $\approx 25.56$. For $m=1$ and $\epsilon/\alpha=3.5$ it is only $\approx 11.37$. 

In any case the amplitude is subject to multiplication by an arbitrary constant (the previous paragraph discusses relative variations), which should be sufficiently small that the asymmetric potential does not dominate the axi-symmetric potential. This can not be done indefinitely if the exponential M instability is really present. In such a case the growth must end by drastically rearranging the disc. This would involve some form of dissipation, probably due to gaseous shocks and star formation. 

An important property of the swinging amplification on an intial spiral as seen in the W function behaviour, is that it rises rapidly at $\zeta=1$ and extends with non-negligible amplitude over a considerable range in $\zeta$. Thus it amplifies over the interesting asymptotic range of the potential. 

We can obtain a plot of the spiral winding in $\zeta$ for given $\nu$ by taking appropriate cuts at constant values of $Re(\Psi_{dr})$. This function is shown in figure (\ref{fig:nuzeta}) and is expressed in equation (\ref{eq:asymrad1}). The cuts for the W and the M components are shown separately in figure (\ref{fig:nuzeta}). We ignore the constant coefficient $T_m(\pi/2)$.

To use this figure one can choose a given $\nu$, which then `recurs' in $\zeta$ (that is increasing time and decreasing radius). We see this recurrence several times corresponding to different windings. The value $\nu+\pi$ then describes the winding of the second arm as a function of $\zeta$. In the range $\nu\rightarrow \nu+ 2\pi$ there are actually two possible pairs shown, but only one pair is relevant depending on the chosen $\nu$. The number of windings are finite for the W component because of the rapid dissipation, while any potential cut gives windings for the M component until the exponential explosion. 
Moving vertically in the figure is equivalent to moving in $\phi$. The negative slope part of the arm reflects approximately the swinging of equation (\ref{eq:dnu}), since near the maximum a cut at constant potential is nearly a cut at constant phase (\ref{eq:evolvpot}). 

It is interesting to note  that the growth always begins first at small radius and then propagates outwards in every case. We find that by reducing the winding angle (increasing the pitch angle) the arms dissipate more rapidly in $\zeta$. At $\epsilon/\alpha=2.5$ the arms begin to dissipate by $\zeta=\approx 4$ rather than $\approx 6$ in the case illustrated in figure (\ref{fig:nuzeta}). The amplified arms are thus more concentrated in time and space. The amplification is  reduced in amplitude for single armed trailing spirals with $m=1$. 

\begin{figure}[t]
\begin{tabular}{cc} 
{\rotatebox{0}{\scalebox{.4}
{\includegraphics{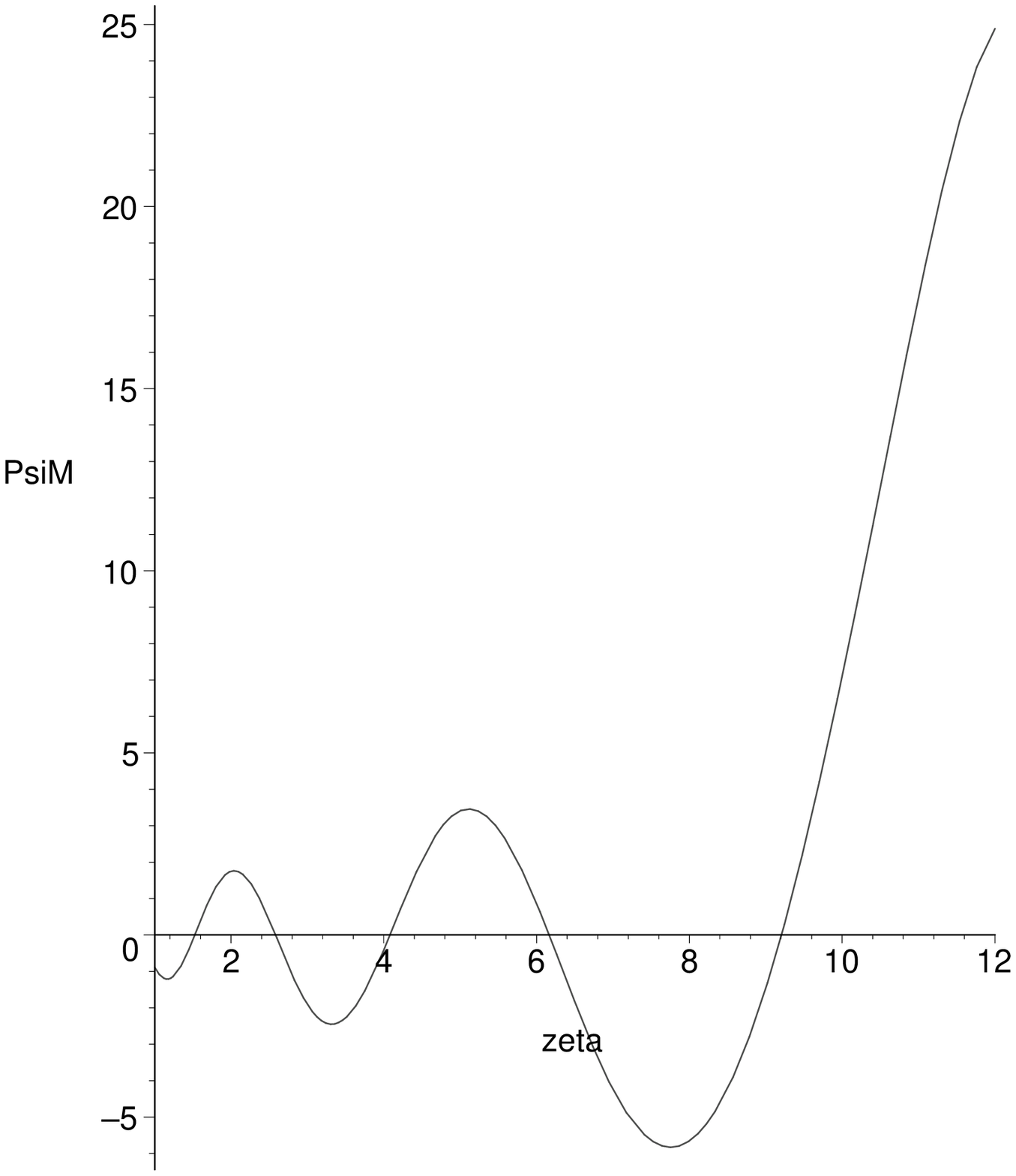}}}}&
{\rotatebox{0}{\scalebox{.4}
{\includegraphics{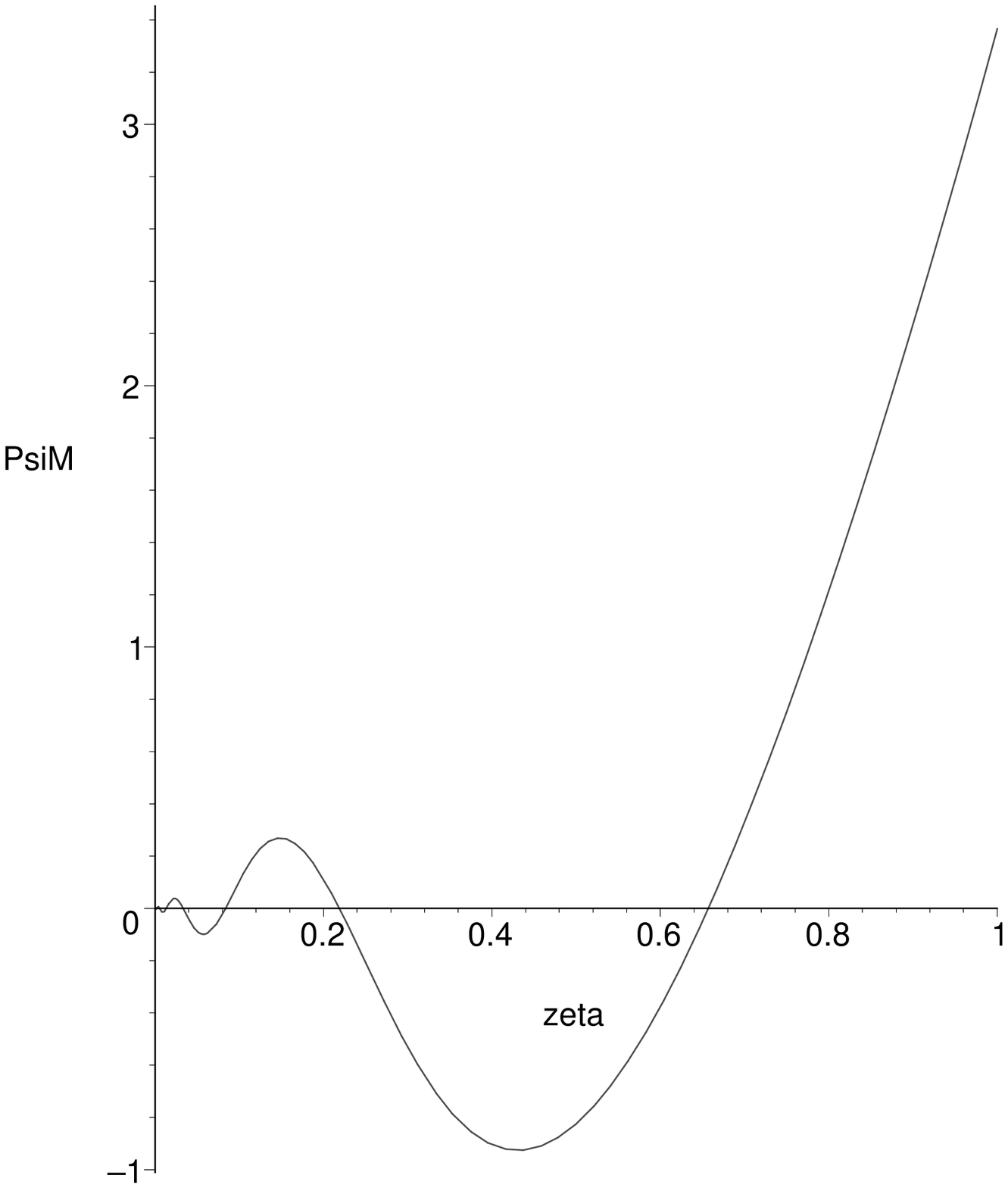}}}}\\
\end{tabular}
\caption{The left figure shows the M potential component for a single armed spiral with $\epsilon/\alpha=3.5$. On the right is the same component of a leading single armed spiral with $\epsilon/\alpha=-3.5$ The instability is much more rapid for the leading spiral.   }    
\label{fig:epsilonm1}
\end{figure}

The real part of the W component of the potential has much the same form for negative $\epsilon/\alpha$, that is for leading spirals. However its amplitude is reduced by many orders of magnitude. The M component is  exponentially unstable much earlier for leading spirals. This is illustrated in  figure (\ref{fig:epsilonm1}) for  spirals with $\epsilon/\alpha=\pm 3.5$ and $m=1$.

\section{Disc Boundary Condition }

The boundary condition (\ref{eq:discsigma}) can only be satisfied at discrete values of $\zeta$, unless there is a collisional asymmetric gas component $\sigma_{gr}$ present in the disc. Such values would approach self-consistent normal modes as closely as possible in the complete self-similar disc. Otherwise a modal analysis would have to be abandoned in favour of integrals over the base functions, preferably using spiral arm based coordinates (e.g. \cite{Harxiv2011}).

In this paper we avoid this analysis (although it is quite feasible) by adding  gas that is after all present in real galactic discs. Assuming a gas contribution converts the condition (\ref{eq:discsigma}) to 
\be
2\pi G(\sigma_{gr}+\sigma_{dr})=-\frac{1}{r}(\partial_\theta\Psi_{dr})|_{\theta=\pi/2},\label{eq:gas+disc}
\ee
where the expression (\ref{eq:collsigmapot}) for $\sigma_{dr}$ applies. The gas surface density must be compatible with the isothermality of the disc, including its spiral structure. We therefore assume that it takes the isothermal self-similar form 
\be
\sigma_{gr}=\Sigma_{gr}(\kappa,\zeta)/(\delta r),
\ee
where it is convenient to take $\delta=\alpha/V$. This converts equation (\ref{eq:discsigma}) into a normal mode analysis for the joint system of gas and disc. In general the gas dynamics is complex and involves the galactic magnetic field as well as pressure and gravity. However this complexity goes beyond the simple normal mode model. The present procedure requires the gas to follow an arm, to which potential the gas behaviour is in fact very sensitive.

We recall that the asymmetric gas density $\sigma_{gr}$ in the condition (\ref{eq:gas+disc}) can be negative, so long as its magnitude does not exceed that of the background axi-symmetric density $\sigma_a$.  
  The self-consistent condition (\ref{eq:gas+disc}) now becomes explicitly for a single mode $m$ ($x=\cos{\theta}$)

\be
\Sigma^m_{gr}(\kappa,\zeta)+2\pi \Phi^{(r)}_{do}K^m_{dr}\exp{(-(\frac{\Psi^m_{dr}(0,\kappa,\zeta)}{\Phi^{(r)}_{do}}))}
=\frac{\alpha/V}{2\pi G }\partial_x\Psi^m_{dr}(x,\kappa,\zeta)|_{x=0}.\label{eq:bcexplicit}
\ee
The potential and its derivative at the disc follow from equation (\ref{eq:Psid}), where the real part of the potential is required. When $\zeta$ is small we have the initial logarithmic spiral and ${\cal R}_m$ may be taken as $1$ in equation (\ref{eq:Psid}). This case has been discussed in reference \cite{Harxiv2012}, but it is of less interest than the evolving arms to be considered here. In the regime of evolution $\zeta\ge 1$ the relevant solution is given by the real part of equation (\ref{eq:evolvpot}), as discussed above.  

Unlike the situation described in (\cite{Harxiv2012}), the solution for $T_m(x)$ given in (\ref{eq:initspiral}) can take a simple form. In the earlier work we anticipated the possibility of using the term in $Q^m_{im\epsilon}$ to cancel an infinity at the poles due to an axi-symmetric disc-halo potential. In the absence of such an infinity we may simply take $C_{2m}=0$ and impose the desired symmetry according to $P^m_{im\epsilon}(-x)=P^m_{im\epsilon}(x)$. 


The form of the spiral potential in our regime of interest for the evolving initially given structure is that of equation (\ref{eq:evolvpot}). Since  $\kappa\equiv \nu+\zeta$, the various functions are most instructively regarded as a function of $\nu$ and $\zeta$. We consider as our example the tightly wound two-armed case with $m=2$ and $\epsilon/\alpha=3.5$. This gives the `modal potential' explicitly at the disc as
\be
\psi_{dr}(0)=\frac{|C_{12}||P^{(2)}_{7i}|_0}{\Phi^{(r)}_{do}}z^\lambda e^{-z/2}\cos{(2(\nu+\zeta/2+\frac{7}{4}\ln{\zeta})+\phi^{(2)}_P(0)+\phi_{12})},\label{eq:modpot}
\ee
where  $z\equiv 2\sqrt{3}\zeta$ and $\lambda\equiv 7\sqrt{3}/2$.
The phase of the associated Legendre function is $\phi^{(2)}_P(0)\approx -0.8915$ and its modulus is $|P^{(2)}_{7i}|_0\approx 4.449\times 10^5$. We have defined $\psi_{dr}\equiv \Psi^{(2)}_{dr}/\Phi^{(r)}_{do}$. The value of the amplification factor $z^\lambda e^{-z/2}$ at the maximum, is $\approx 8.64\times 10^3$. 

We are free to choose the arbitrary phase $\phi_{12}=-\phi^{(2)}_P(0)$ to simplify the argument of the cosine in the expression for $\psi_{dr}$. This choice of phase  renders the potential negative on the arm. If moreover as an example we take $\psi_{dr}=10^{-3}$ and the cosine function times the amplification factor  to be $1$, then the corresponding $\nu(\zeta)$ curves are shown at left in figure (\ref{fig:nuzeta}). This is because the amplification factor times the cosine function gives $W$ in this regime, and figure (\ref{fig:nuzeta}) is plotted for $Re(W)=1$. 
Numerically the chosen value of $\psi_{dr}$, together with the value of modulus of the Legendre function, imply that we have set $|C_{12}|/\Phi^{(r)}_{do}\approx 2.25\times 10^{-9}$. 

The straight line in figure (\ref{fig:nuzeta}) that passes through the point $\nu=10$, $\zeta\approx 6.458375$, has the approximate equation 
\be
2(\nu+\frac{\zeta}{2}+\frac{7}{4}\ln{\zeta})\approx 21\frac{\pi}{2}-4.7415\times 10^{-4},\label{eq:10curve}
\ee
and we can use this to evaluate the potential and its derivative at this point on an evolving arm that was initially characterized by this value of $\nu$. Substituting relation (\ref{eq:10curve}) into equation (\ref{eq:bcexplicit}) then gives the transient surface density of the gas on the arm. 

To proceed in a less encumbered way we write condition (\ref{eq:bcexplicit}) in the compressed form
\be
S_g(\nu,\zeta)=q\partial_x\psi_{dr}|_{(x=0)}-e^{-\psi_{dr}},\label{eq:Sg}
\ee
where
\bea
S_g&\equiv& \frac{\Sigma^{(2)}_{gr}}{2\pi \Phi^{(r)}_{do}K^{(2)}_{dr}},\nonumber\\
q&\equiv& \frac{\alpha/V}{4\pi^2 GK^{(2)}_{dr}}.
\eea
Thus $S_g$ is the gas density essentially in units of the collisionless density. The parameter $q$ is a ratio of the characteristic dynamic scale to a gravitational scale namely $\alpha L_G/V$, where the gravitational scale $L_G\equiv 1/(4\pi^2 GK^{(2)}_{dr})$. The ratio is similar in spirit to the Toomre $Q$ value, if $\alpha/(2\pi)$ is identified with the epicyclic frequency $\kappa=\sqrt{2}\Omega$,  $V/(2\pi)$ replaces the velocity dispersion, and $KV^2$ replaces the surface density.

The derivative of the potential takes the form at the disc 
\be
\partial_x\psi_{dr}|_0 =\frac{|C_{12}||dP^{(2)}_{7i}/dx|_0}{\Phi^{(r)}_{do}}z^\lambda e^{-z/2}\cos{(2(\nu+\zeta/2+\frac{7}{4}\ln{\zeta})+\phi^{(2)}_{DP}+\phi_{12})},
\ee
where the modulus $|dP^{(2)}_{7i}/dx|_0\approx 3.24\times 10^6$, and the phase $\phi^{(2)}_{DP}\approx -0.9576$ so that$ \phi^{(2)}_{DP}+\phi_{12}\approx -0.062$. 

We calculate the gas density on this arm ($\nu=10$ initially) at $\zeta=6.4584$ by inserting $\nu(\zeta)$ from equation (\ref{eq:10curve}) into equation (\ref{eq:Sg}). Some  calculation using our various numbers gives $S_g\approx- 0.959q-1$ on this arm at the chosen $\zeta$. Thus there is a deficiency in the gas density at this $\zeta$.

\begin{figure}[]
\begin{tabular}{cc} 
{\rotatebox{0}{\scalebox{.6} 
{\includegraphics{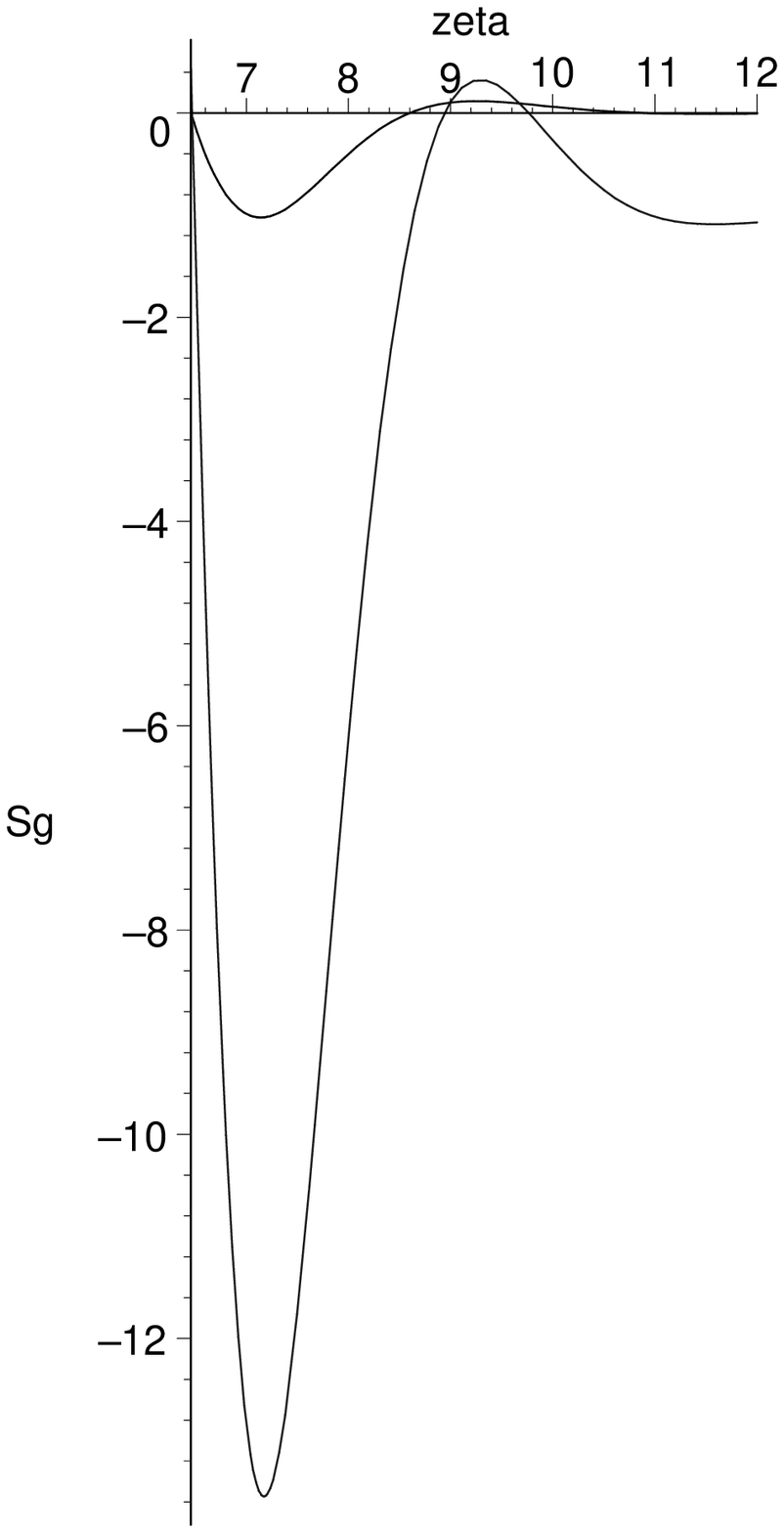}}}}&
{\rotatebox{0}{\scalebox{.4}
{\includegraphics{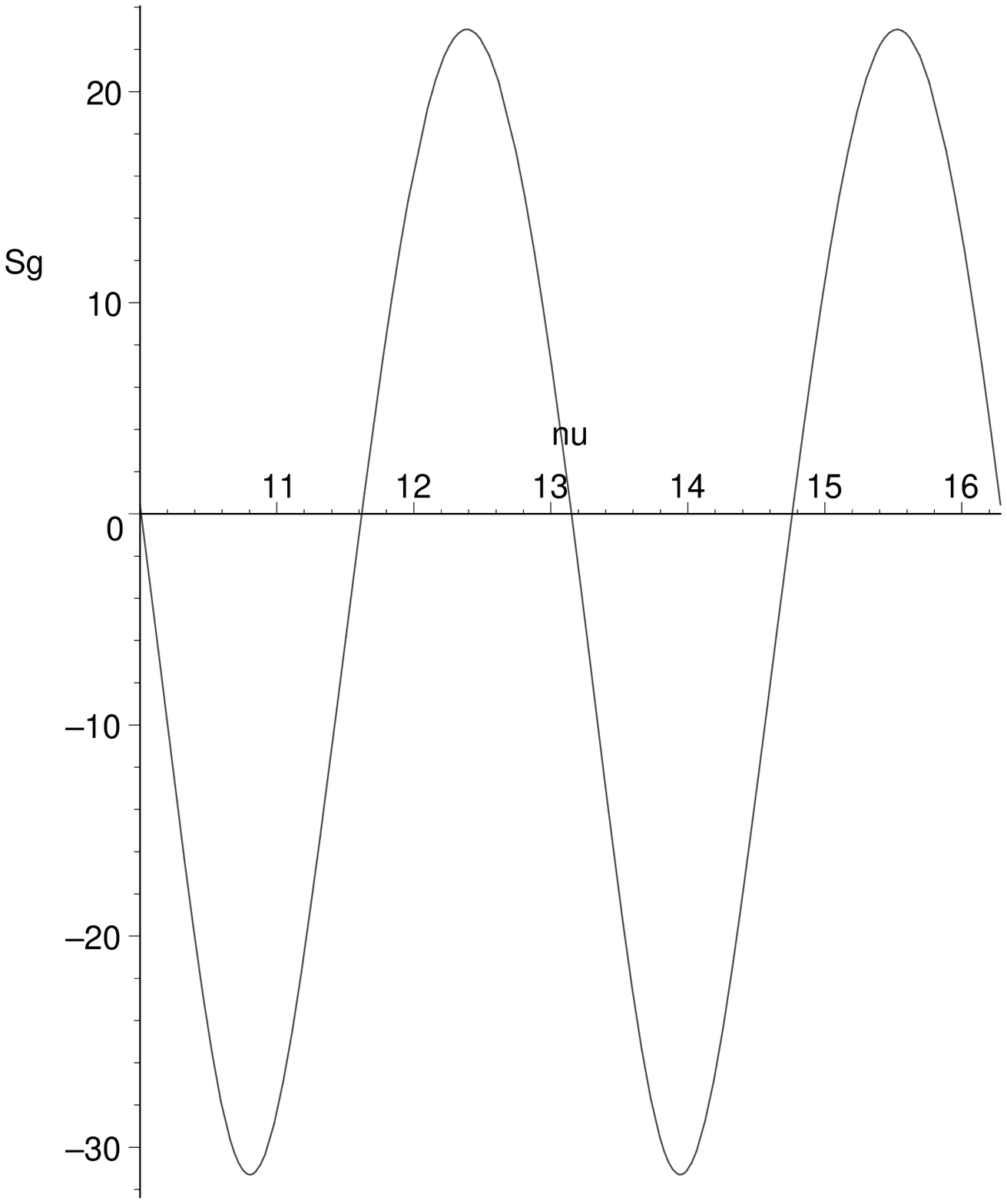}}}}\\
\end{tabular}
\caption{On the left the gas density that satisfies the boundary condition (\ref{eq:Sg}) for $\nu=10$, $m=2$ and $\epsilon/\alpha=3.5$ is shown as a function of $\zeta$ by the large curve. The small curve shows $\psi_{dr}$. In each case we start with the rapid winding of the arm at $\zeta=6.458$. As $\zeta$ increases with decreasing $r$ at a fixed time, the plot shows the gas density falling steeply to a minimum from an excess (not shown) on the outside of the arm, then rising to modest positive values as we leave the arm on the inside, and finally attaining a constant negative value inside the arm. The figure at the right shows the variation of the gas density in increasing angle (that is with increasing $\nu$ at fixed radius) at a fixed $\zeta=6.458$. In each case $q=1.5$. The peaks are on the outside of the arms. }    
\label{fig:gas}
\end{figure}

The variation of the gas density with $\zeta$ that follows from equation (\ref{eq:Sg}) is shown by the large curve on the left in figure (\ref{fig:gas}) for an arm with an initial $\nu=10$. The calculation follows the line passing through the point $(\zeta=6.4584,\nu=10)$ in figure (\ref{fig:nuzeta}), which is given by equation (\ref{eq:10curve})). This is the region where the initial arm is rapidly winding. 

We note that, starting from a large excess on the outside of the arm (small $\zeta$-not fully shown in order to display the potential), the gas density drops to strongly negative values as one crosses the arm towards the inside(larger $\zeta$ at fixed time). Ultimately, leaving the arm on the inside, the gas density rises to a modest positive maximum and finally finishes as a negative constant where the resonant function drops to zero. On the same dimensionless scale the smaller curve gives the variation of $\psi_{dr}$ with $\zeta$. The gas density minimum is slightly inward of the potential minimum in the arm.

It should be noted that the amplitude of the variation in the gas density is very sensitive to the amplitude of the potential. Increasing this by only one factor of ten renders the swing in the gas density quite unreasonable. Such strong dynamic behaviour would require a full  treatment of shocks and magneto-gas-dynamics. Given $m$ and the winding angle $\epsilon/\alpha$, the only real free parameters are this amplitude and $q$, the analogue of the Toomre $Q$. Our examples should thus be representative of real behaviour.
   
The right-hand side of the figure gives the angular dependence of the gas density at a fixed $\zeta$ and radius ($3.5\ln{\alpha r/V}=10$). The two peaks are separated by two slighter deeper troughs. Since the angle increases in the sense of rotation, one encounters the peaks on the outside of the arms.

\section{Conclusions}

This paper has studied finite spiral structure in isothermal (constant rotational velocity) discs. The collisionless particles that comprise these spiral `arms' are described by an isothermal distribution function in the locally rotating frame. The particles, the arm,and indeed the potential, therefore comove.

The presence of gas is a necessary feature. We calculate its distribution based solely on the requirement that the thin disc self-consistent condition (\ref{eq:Sg}) be satisfied. The gas peaks and valleys are associated respectively with the potential peaks and valleys, so in this approximation it is also comoving. The gas density is dramatically higher on the outside of an arm.  

These structures are not steady and we found the time dependent potential that describes their evolution. For each  `mode' (we use this expression for the potential that satisfies equation (\ref{eq:Sg}) for each $m$), this potential possesses two growth behaviours as shown in figure (\ref{fig:potcomps}). In the Whittaker W swing amplification studied mainly in this paper, the evolution of an initial arm is followed in time and space. This arm is first amplified as it `swings' in pitch angle and finally dissipates exponentially. The `swinging' and the amplification function are illustrated in figures (\ref{fig:windingspirals}) and (\ref{fig:nuzeta}). We use a winding angle of $74^\circ$ (that is a pitch angle of $16^\circ$) for a two-armed spiral. This is tight winding. The argument is readily extended to other winding angles and normal modes ($m$). 

The disturbance in the gas density that follows  by requiring the self-consistent condition (\ref{eq:Sg}) to be satisfied everywhere on the disc, is shown in figure(\ref{fig:gas}) for an initial arm undergoing swing amplification. The gas density is increased on the outside of the arm, declines strongly on the potential arm, and rises slightly inside of the arm. It finally attains a constant negative value inside the arm as the potential declines exponentially. At a fixed radius the same sequence is encountered in time as the arm crosses that radius. Such a sequence must be restarted in some undefined fashion. The variation in angle at a fixed value of the self-similar variable zeta is also shown in that figure.

The amplitude of the variation in the gas density becomes exaggerated relative to any reasonable background, if the ratio of the asymmetric potential amplitude is greater than about $10^{-3}$ times the radial potential amplitude (essentially $\Phi^{(r)}_{do}$). In reality, stronger disturbances would require solving for the actual magneto-gas-dynamics in the asymmetric potential.

The `self-excited' instability of figures (\ref{fig:potcomps}, \ref{fig:epsilonm1}) seems to duplicate behaviour that which occurs in simulations of an isolated disc (\cite{JS2011}). Its origin in a central singularity renders its physicality suspect. However it may indicate the presence of a genuine instability, which is due to the interchange of leading and trailing waves across the centre of the system. All that one can say definitely is that the amplification of both leading and trailing waves originates at the centre.

Taking it at face value, the exponentially growing disturbance must be stabilized in some dissipative fashion, but we can not follow this development in general beyond a maximum value of $\zeta=Vt/r$. For completeness we see from figure (\ref{fig:epsilonm1}) that the instability is slightly less rapid for a one-armed spiral than that of a two-armed spiral of the same winding angle. However it is much faster for the one-armed leading spiral of the same winding angle.
 
In the case of a larger number of arms, such as $m=3$, one finds (not shown here) the self-excitation to be similarly strong for the same winding angle. The swing amplification of an initial arm is of much larger amplitude however, unless the winding angle is reduced.

On the whole this analysis suggests that galactic spirality is transient and either continually self-excited after each cycle or re-excited by external influences. It has a dramatic effect on the gas distribution in the disc, which is in turn essential to the continuous spirality.

\section{Acknowledgements}

Queen's University at Kingston is to be thanked for their partial support of this research. Judith Irwin is to be thanked for helpful comments. Several referees have struggled to clarify the arguments without really finding them to be erroneous. I am grateful for their efforts.

\enlargethispage*{1000pt}

\label{lastpage}

 \end{document}